# Method to deterministically generate large-amplitude optical cat states

Zheng-Hong Li,[1,2,*] Fei Yu,[1] Zhen-Ya Li,[1] M. Al-Amri,[3,4,5,*] and M. Suhail Zubairy[3,*]

*1 Institute for Quantum Science and Technology and Department of Physics, Shanghai University, Shanghai 200444, China*

*2 Shanghai Key Laboratory of High Temperature Superconductors, Shanghai University, Shanghai 200444, China*

*3 Institute for Quantum Science and Engineering (IQSE) and Department of Physics and Astronomy, Texas A&M University, College Station, Texas 77843-4242, USA*

*4 NCQOQI, KACST, P.O.Box 6086, Riyadh 11442, Saudi Arabia*

*5 Institute of Quantum Technologies and Advanced Computing, KACST, Riyadh 11442, Saudi Arabia*

**Abstract**

Cat states, as an important resource in the study of macroscopic quantum superposition and quantum information applications, have garnered widespread attention. To date, preparing large-sized optical cat states has remained challenging. We demonstrate that, by utilizing interaction-free measurement and the quantum Zeno effect, even a fragile quantum microscopic system can deterministically control and become entangled with strong light fields, thereby generating large-amplitude optical cat states. During the entire preparation process, our method ensures that the microscopic system functions within a weak field environment, so that its quantum property can be protected. Furthermore, we show that the preparation of cat states is possible even when the quantum microsystem suffers from significant photon loss, provided that optical losses from classical devices are kept low, which implies that the fidelity of the cat state can be enhanced by improvements to and the perfection of the classical optical system.

## I. Introduction

Schrödinger's gedanken experiment involving a cat in a superposition of dead and alive states vividly demonstrates the magical picture of quantum superposition on a macro scale [1]. In modern physics, this cat state (CS) is usually represented by the superposition of two distinct coherent states $|\pm\alpha\rangle$, which can generally be considered orthogonal when $|\alpha| \geq 2$. With the increase of amplitude $|\alpha|$, CS gets closer to the macroscopic superposition, making it an important resource for the study of macroscopic quantum phenomena. CS is not only attractive from a fundamental point of view [2,3], but also valuable for applications including quantum teleportation [4-6], quantum computing [7-11], quantum error correction [12-17] and quantum metrology [18-24]. After decades of efforts, CS is now being generated on various platforms [3,25 -28]. However, even for the best experimental results in the optical domain so far $|\alpha|$ remains less than 2 [29-37], and it remains a challenge to increase the value of $|\alpha|$ based on existing methods. For example, the widely studied photon subtraction method [29-33,38,39], which is a probabilistic method based on post-selection, has low probability of success for generating large-amplitude CS. Another approach that has attracted a lot of attention is the synthesis method [36,40,41], which uses a pre-prepared CS as a seed to grow large-amplitude CS. However, this method is still limited by the amplitude of the pre-prepared CS. Furthermore, a deterministic method that based on light-matter interaction [42] has recently been demonstrated experimentally [37]. In this method, CS is generated by direct interaction between an incident coherent light pulse and a single-side cavity containing a three-level atom [37,42-45]. Its basic idea is to use normal-mode splitting [43] to control the phase of the reflected coherent light field. In the experiment, the amplitude of the output CS is $|\alpha| \approx 1.4$ [37]. Unfortunately, this method is limited in preparing an arbitrarily large $|\alpha|$ CS for a given optical pulse length, due to the fragility of the atom-cavity coupling system when interacting with a strong light field. As the external field strengthens, a single atom ultimately cannot prevent the optical field from entering the cavity, resulting in the reflected field carrying a $\pi$ phase shift regardless of the atomic state, similar to what occurs with empty cavity reflection.

Needless to say, optical field is an excellent medium for information transmission. Considering the practical application requirements that involve integrating diverse quantum systems, as well as for advanced sensing and imaging techniques, the generation of optical CS with large amplitudes and customizable propagation properties becomes not only necessary but also of significant value. This endeavor represents a pioneering frontier in the realm of quantum optics, bearing profound implications for both the theoretical understanding and the advancement of practical quantum technology [46].

In this article, we propose a deterministic method to generate flying optical CS whose amplitude can be arbitrarily large, provided that optical losses from classical optical devices remain low. The key step of our scheme is to entangle a strong coherent light field with a microscopic quantum system prepared in a superposition state initially. After the entanglement, the optical CS is obtained by performing projection measurement on the quantum microsystem. In response to the situation that a quantum microsystem is typically very fragile when facing strong light fields, our solution is to utilize interaction-free measurement based on quantum Zeno effect [47-49], which is considered a powerful non-invasive measurement method that can protect vulnerable samples. In this work, we further apply it to protect the quantum properties of the microscopic system from destruction by strong fields. We replace the direct interaction between the strong light field and the microsystem with a sequence of interactions, each with only a small fraction of photons actually interacting with the microsystem. This ensures that the quantum microsystem always works in a weak field environment, making one of the main advantages of this work. Eliminating the negative influence of strong fields gives liberty of applying this method in a wide variety of systems and provides the potential for future applications. More importantly, this indicates that our scheme allows quantum microscopic systems to manipulate macroscopic counterparts that are challenging to control directly. In addition, our method has another important advantage—it can reduce the requirements of quantum microscopic systems. For illustration, we employ the single-side cavity-atom coupling model [37] as our specific quantum microscopic system. Our simulation shows that our scheme becomes less and less sensitive to both atomic spontaneous emission and detuning between the atom and the cavity as the number of interactions increases, when other classical optical devices are ideal. This enables our method to reduce the impact of optical losses originating from quantum device (referred to as Type I light loss), with the main limitation of our method arising from the optical losses introduced by classical optical devices (referred to as Type II light loss). Finally, it is also worth mentioning that here we mainly focus on the optical platform, but our approach can also be extended to other platforms such as superconducting microwave resonator [50,51].

## II. Results and Discussion

## A. General multiple reflection scheme

First, we focus on explaining the principle of our method. More detailed experimental protocols will be discussed later. We start with the chain Mach-Zehnder interferometer shown in Fig. 1a, which is a commonly used optical structure for interaction-free measurement [52-57]. Compared to conventional interaction-free measurement studies, our scheme has two distinct features, although these are commonly used and even necessary for CS preparation. Firstly, we use a coherent light source (S) instead of a single photon source. Secondly, conventional interaction-free measurement typically assume that an object has two functions: absorbing photons or allowing them to pass through. However, our method introduces an additional function for the object, which is to apply a π phase shift to the light field passing through it. In details, as shown in Fig.1a, $\mathrm{MR}$ stands for normal mirror, and $\mathrm{BS}$ stands for beam-splitter. The interference occurs between the light fields in Zones 0 and 1, which are located on the lower and upper sides of $\mathrm{BS}$, respectively, separated by a dotted line. Assuming that $a_z^\dagger$ $(z=0,1)$ represents the creation operator of the light field in Zone $z$, the function of $\mathrm{BS}$ can be described by $a_0^\dagger \to a_0^\dagger \cos\theta + a_1^\dagger \sin\theta$ and $a_1^\dagger \to a_1^\dagger \cos\theta - a_0^\dagger \sin\theta$ [49], where $\cos^2\theta$ represents the reflectivity of $\mathrm{BS}$. As for the object, it has three states, $|\mathrm{pass}\rangle$, $|\mathrm{block}\rangle$, and $|\mathrm{phase}\rangle$. Firstly, we discuss the state $|\mathrm{pass}\rangle$, where the object is transparent, thereby causing optical interference to occur continuously in Zones 0 and 1. Assuming the initial state of the light field is such that the field in Zone 0 is in the coherent state $|\alpha\rangle$ and the field in Zone 1 is in the vacuum state $|0\rangle$, it is represented as $|\alpha,0\rangle = \exp(\alpha a_0^\dagger - \alpha^* a_0)|0,0\rangle$. After passing through $N$ number of $\mathrm{BSs}$, it becomes $|\alpha\cos N\theta, \alpha\sin N\theta\rangle$ [58]. Apparently, when $\theta = \pi/2N$, we have $|0,\alpha\rangle$. While when $\theta = \pi/N$, we have $|-\alpha,0\rangle$. Secondly, we discuss the state $|\mathrm{block}\rangle$. In this case, any photon passing through the object is absorbed. Notice that the object and the light field interact many times. To take into account all absorbed photons, we write the state of the light field as $|C_0, C_1\rangle|l_1, l_2, \dots l_{N-1}\rangle$, where $|C_{z=0,1}\rangle$ represents photon state in Zone $z$ and $|l_1, l_2, \dots l_{N-1}\rangle$ represents the photon absorbed each time. After the light pulse passing through the first $\mathrm{BS}$, the initial state $|\alpha,0\rangle|0,0,\dots 0\rangle$ evolves to $|\alpha\cos\theta, \alpha\sin\theta\rangle|0,0,\dots 0\rangle$, and it further becomes $|\alpha\cos\theta, 0\rangle|\alpha\sin\theta, 0, \dots 0\rangle$ due to the absorption of the object. Subsequently, the photons remaining inside the interferometers pass through the next $\mathrm{BS}$ and interact with the object again. Such process repeats $N$times. After $N$ numbers of $\mathrm{BSs}$, the photon state is $|\alpha\cos^N\theta, \alpha\cos^{N-1}\theta\sin\theta\rangle \prod_{n=1}^{N-1}|\alpha\cos^{n-1}\theta\sin\theta\rangle$. Note that when $\theta = \pi/N$, as $N\to\infty$, we have $\cos^N\theta \to 1$ and $\sin\theta \to 0$. Therefore, as long as $|\alpha|^2$ is finite, the light field can be frozen in its initial state $|\alpha,0\rangle|0,0,\dots 0\rangle$. This also means that the actual number of photons absorbed by the object during the entire process is close to zero. The object is spared from being directly exposed to light field due to the interruption of the interference process, which can be considered the quantum Zeno effect caused by photon loss. Besides the state $|\mathrm{block}\rangle$, the state $|\mathrm{phase}\rangle$ can also maintain the light field in its initial state by applying a π phase shift $(a_1^\dagger \to -a_1^\dagger)$ to the photons passing through it. When the photons encounter an odd number of $\mathrm{BS}$, the state becomes $|\alpha\cos\theta, \alpha\sin\theta\rangle$, while after passing through an even number of $\mathrm{BS}$, it becomes $|\alpha, 0\rangle$. It's important to note that the maximum intensity of photons actually passing through the object during each interaction is $|\alpha\sin\theta|^2$. When $\theta = \pi/N$, this value tends to zero as $N$ increases, suggesting the freezing of the light field. Consequently, for both states $|\mathrm{block}\rangle$ and $|\mathrm{phase}\rangle$, the object works in a weak field environment.

Above, we have introduced the three states of the object and the corresponding dynamic evolution behavior of the light field. It is easy to see that if an object is initially in a superposition state of $(|\mathrm{pass}\rangle + |\mathrm{block}\rangle)/\sqrt{2}$ or $(|\mathrm{pass}\rangle + |\mathrm{phase}\rangle)/\sqrt{2}$, the object can be entangled with the coherent light field when $\theta = \pi/N$, i.e., $(|-\alpha, 0\rangle|\mathrm{pass}\rangle + |\alpha, 0\rangle|\mathrm{block/phase}\rangle)/\sqrt{2}$. After measuring the object with $(|\mathrm{pass}\rangle \pm |\mathrm{block/phase}\rangle)/\sqrt{2}$, CS can be deterministically created. However, whether it turns out to be an odd or even CS is random, each with a $50\%$ probability of occurrence. The even CS occurs when the measurement result is $(|\mathrm{pass}\rangle + |\mathrm{block/phase}\rangle)/\sqrt{2}$. As for the other measurement result, the odd CS is obtained. In addition, it is worth mentioning that our method can also realize the entangled coherent state $(|\alpha, 0\rangle + |0, \alpha\rangle)/\sqrt{2}$ [20,21]. To do so, we just need to set $\theta = \pi/2N$. Given that the optical field is reflected and interacts with the quantum object many times, we refer to the above method as the multiple reflection scheme in the following discussion.

### B. Multiple reflection scheme using single-side cavity-atom coupling system

The above is the basic idea of our method. The next important issue is what kind of real quantum microscopic system can be used as an object for our scheme. We have shown that it is optional whether the object absorbs photons or provides a π phase. The difference is that the latter way does not introduce any photon loss. In theory, it has better fidelity. Nevertheless, given that photon loss due to device imperfection is inevitable in real systems, the former mechanism is also critical. In fact, our later numerical calculations will show that such mechanism can even mitigate the impact of optical losses caused by the quantum system, indicating that the optical loss of the quantum microsystem no longer serves as the decisive factor when choosing a microsystem for our multiple reflection scheme. Microsystems capable of implementing optical path blocking, such as Rydberg blockade [59-61], photon blockade [62], and so on [58,63] , remain viable candidates for applying our method. Regardless of the system type, our scheme allows the microsystem to function properly within a weak field environment. This represents the primary contribution of our work, enabling our scheme to be utilized in any given experimental environment and offering potential for future practical application. In the following, we will analyze a detailed experimental scheme based on a specific microscopic system in order to study the performance of our method under realistic conditions.

For illustrative purposes only, we adopt the atom-single side cavity system [37] as our quantum microsystem. As shown in Fig.1b, a single atom is embedded in a single-side cavity $\mathrm{SSC}_1$, which is constituted by two facing mirrors $\mathrm{CM_R}$ and $\mathrm{CM_T}$. Ideally, $\mathrm{CM_T}$ is assumed to have perfect reflection, but $\mathrm{CM_R}$ is allowed for in- and outcoupling of light. The energy level configuration of the atom is shown in Fig.1c. The transition between levels $|\uparrow\rangle$ and $|e\rangle$ is coupled by the cavity mode. When the atom is in $|\uparrow\rangle$, the weak coherent light pulse, resonant with the cavity, is prevented from entering the cavity due to normal-mode splitting, which results in reflection without a π phase shift. As for the transition between $|\downarrow\rangle$ and $|\uparrow\rangle$, it is decoupled from the cavity mode due to large detuning. Therefore, when the atom is in $|\downarrow\rangle$, the cavity can be treated as an empty cavity. The incident pulse enters the cavity and reflects back but with a π phase. This atom-cavity coupling model has been directly used to generate CS [37,42]. After a single reflection (Hereinafter we call it the single reflection scheme), the atom-field entanglement $(|\alpha\rangle|\uparrow\rangle + |-\alpha\rangle|\downarrow\rangle)/\sqrt{2}$ can be generated. Unfortunately, we will later explain why the method becomes ineffective when the incident light field is strong, but our solution can address this problem.

Based on the atom-cavity system, our detailed experimental scheme is presented in Fig.1b. In order to implement multiple interactions between the atom and light field, we do not directly use the chain interferometer structure in Fig.1a, but rather its equivalent Michelson interferometer. In the figure, SM stands for switchable mirror (In the experiment, it can be realized by a fiber switch [64] or Pockels cell [65-68]), which is transparent only at the beginning and end of preparation. At all other times, it is just a normal mirror, forcing the light pulse to bounce back and forth inside the interferometer for $M$ cycles. One cycle is defined by a pulse starting at SM, going through BS twice, and returning to SM. Because the optical field passes through the BS twice during each interaction with the atom, we adjust the parameter of BS to $\theta = \pi/2M$. In addition, $SSC_0$ is an empty single-side cavity. Except for the absence of atom, everything else is the same as $SSC_1$. C stands for optical circulator. PS stands for phase shifter. Unlike other components, the use of PS is only for the convenience of calculation and discussion. Its removal does not have a substantial impact on our method. We assume that PS adds a π phase shift to the light field only as it propagates from BS to SSC. Its function can be described as $a_z^\dagger \rightarrow -a_z^\dagger$, where we still set $a_z^\dagger$ $(z = 0,1)$ as the creation operator of the light field in Zone $z$.

Initially, the light source emits a coherent pulse into the interferometer, while the light field in Zone 1 is in a vacuum state. As for the atom, it is prepared in a superposition state. Accordingly, the initial state of the entire system is $|\alpha, 0\rangle(|\uparrow\rangle + |\downarrow\rangle)/\sqrt{2}$. After $m$ number of cycles, the wave-function of the whole system becomes (See the "System wave-function calculation" subsection in the Methods)

$$\left|\psi^{(2m)}\right\rangle = \frac{1}{\sqrt{2}}(|\alpha, 0\rangle|\uparrow\rangle + |\alpha\cos 2\, m\theta_M, \alpha\sin 2\, m\theta_M\rangle|\downarrow\rangle). \tag{1}$$

The dynamic evolution process of the system can be briefly described as follows. Due to the presence of the empty cavity $SSC_0$, when the atom is in $|\downarrow\rangle$, light field has no phase difference in Zones 0 and 1. Interference continues to occur, thereby generating $|-\alpha\rangle$. On the contrary, when the atom is in $|\uparrow\rangle$, the light field is frozen in $|\alpha\rangle$ due to the phase difference between the two Zones. Consequently, after $M$ cycles, we have the light-atom entangled state $(|\alpha, 0\rangle|\uparrow\rangle + |-\alpha, 0\rangle|\downarrow\rangle)/\sqrt{2}$. We note that the light field is output from the $SM_0$ end and no photons appear at $SM_1$ side. After measuring the atom, the optical CS is deterministically prepared.

So far, we have only focused on the ideal case. In the following, we analyze the performance of our multiple reflection scheme for non-ideal situation. We show that our scheme is highly durable when it comes to parameters such as atomic spontaneous emission decay and atom-cavity detuning. The reason for this is the quantum Zeno effect caused by light absorption.

### C. Experimental parameter analysis of multiple reflection scheme based on single-side cavity-atom coupling system

In the upcoming discussion, we utilize numerical simulations to analyze the impact of various practical parameters, including atomic, cavity parameters, the number of cycles, and the light loss rates of classical optical devices. We employ fidelity and cattiness [69-71] as metrics to assess the quality of the CS generated by our method. Additionally, recognizing the critical role of light loss in CS preparation [71], the two types of optical losses categorized in the introduction are the focus of our discussion. The first type is caused by the quantum system, specifically the atom-cavity interaction, while the second type solely originates from classical optical devices.

To begin with, we briefly outline the definitions of each parameter. Regarding the practical atom-cavity system ($\mathrm{SSC}_1$), the incident light field is not only reflected, but also transmitted and scattered. To evaluate these effects, we set that $2\gamma$ and $\omega_{\mathrm{a}}$ as the spontaneous emission decay rate and transition frequency of the atomic transition between $|\mathrm{e}\rangle$ and $|\uparrow\rangle$, respectively. The coupling constant between the cavity mode of frequency $\omega_{\mathrm{c}}$ and the atomic transition is $g$. The atom-cavity detuning is $\Delta = \omega_{\mathrm{a}} - \omega_{\mathrm{c}}$. Moreover, we set $\kappa_{\mathrm{R(T)}}$ as the cavity field decay rate into the external light field on the $\mathrm{CM}_{\mathrm{R(T)}}$ side. Given that the atom is hardly excited in our scheme, as long as the condition of slowly varying light intensities is satisfied [37,45,72], $\mathrm{SSC}_1$ can be well described by the input-output theory [73,74]. Suppose that $|\alpha_{\mathrm{i},1\uparrow}\rangle$ is the incident coherent light field from $\mathrm{CM}_{\mathrm{R1}}$ side when the atom is in $|\uparrow\rangle$. The cavity reflection $|\alpha_{\mathrm{R},1\uparrow}\rangle$ satisfies (See the "Input-output theory for the atom-cavity coupling system" subsection in the Methods)

$$\alpha_{\mathrm{R},1\uparrow} = \left[1 - \frac{2\kappa_{\mathrm{R}}(\mathrm{i}\Delta+\gamma)}{\kappa(\mathrm{i}\Delta+\gamma)+g^2}\right]\alpha_{\mathrm{i},1\uparrow} = \left|\eta_{\mathrm{R},1\uparrow}\right| e^{\mathrm{i}\beta_{\mathrm{R},1\uparrow}}\alpha_{\mathrm{i},1\uparrow}, \tag{2}$$

where $\kappa = \kappa_{\mathrm{R}} + \kappa_{\mathrm{T}}$, $\left|\eta_{\mathrm{R},1\uparrow}\right|^2$ is the reflectivity and $\beta_{\mathrm{R},1\uparrow}$ describes the phase of the reflection. Similarly, for the transmission of the cavity $\left|\alpha_{\mathrm{T},1\uparrow}\right\rangle$, we have

$$\alpha_{\mathrm{T},1\uparrow} = -\frac{2(\mathrm{i}\Delta+\gamma)\sqrt{\kappa_{\mathrm{T}}\kappa_{\mathrm{R}}}}{\kappa(\mathrm{i}\Delta+\gamma)+g^2}\alpha_{\mathrm{i},1\uparrow}. \tag{3}$$

Regarding the scattering field $\left|\alpha_{\mathrm{S},1\uparrow}\right\rangle$ due to the atomic spontaneous emission, we have

$$\alpha_{\mathrm{S},1\uparrow} = \frac{2g\sqrt{\kappa_{\mathrm{R}}\gamma}}{\kappa(\mathrm{i}\Delta+\gamma)+g^2}\alpha_{\mathrm{i},1\uparrow}. \tag{4}$$

As for the situation that the atom is in $|\downarrow\rangle$, we still assume that the atom is completely unaffected by the cavity mode due to the large detuning. Therefore, $\mathrm{SSC}_1$ in such case can be treated the same as the empty cavity $\mathrm{SSC}_0$. By setting $g = 0$ in Eqs. (2)-(4), we can immediately obtain the corresponding reflection and transmission. As for the scattering light field, it is 0.

Before we continue our parameter analysis, let us make some necessary discussions about Eqs. (2)-(4). They constitute the mathematical foundation of the single reflection model, and seem unrelated to the intensity of the input light $|\alpha|^2$. It is true that the single reflection model allows for a certain degree of flexibility in $|\alpha|^2$. However, to derive Eqs. (2)-(4), the weak excitation approximation is required (see Eq. (15) in Methods), meaning that the light field inside the cavity cannot be too strong. This leads to the requirement that the cavity parameter $\kappa_{\mathrm{R}}$ not be too large, and similarly, the intensity of the incident field or the peak value of the incident pulse must also be constrained. In fact, when parameters such as $\kappa_{\mathrm{R}}$, $\kappa_{\mathrm{T}}$, $\gamma$, $\Delta$, $g$ and pulse length are fixed, increasing the intensity of the incident light always invalidates Eqs.(2)-(4). Consequently, to guarantee the single reflection model functions properly, further improvement of the system parameters is necessary. Overcoming these limitations is one of the objectives of our work. Our scheme ensures that the single-side cavity-atom coupling system can always work under the appropriate external field intensity. This effectively leads to avoid the difficulties brought about by improvement of the quantum system. From this perspective, our method is not a negation of the single reflection method. The core principle of our approach is to enhance and broaden the operational range of the atom-cavity coupling system. The two methods are complementing each other.

Based on Eqs. (2)-(4), we can numerically simulate the dynamic evolution of the input coherent pulse, and the fidelity of the output in the multiple reflection scheme. We suppose that the target state is $|\psi_{\rm T}\rangle = (|\alpha\rangle|\uparrow\rangle + |-\alpha\rangle|\downarrow\rangle)/\sqrt{2}$, and the final state of the whole system after $M$ cycles is $|\psi_{\rm f}\rangle = (|C_{0\uparrow}\rangle|{\rm loss}_\uparrow\rangle|\uparrow\rangle + |C_{0\downarrow}\rangle|{\rm loss}_\downarrow\rangle|\downarrow\rangle)/\sqrt{2}$ with $|{\rm loss}_{\uparrow(\downarrow)}\rangle = |C_{1\uparrow(\downarrow)}\rangle \otimes \prod_{m=1}^{M}\left|\alpha_{{\rm T},0\uparrow(\downarrow)}^{(m)}\right\rangle\left|\alpha_{{\rm S},0\uparrow(\downarrow)}^{(m)}\right\rangle\left|\alpha_{{\rm T},1\uparrow(\downarrow)}^{(m)}\right\rangle\left|\alpha_{{\rm S},1\uparrow(\downarrow)}^{(m)}\right\rangle$. Here state $|C_{z\uparrow(\downarrow)}\rangle$ $(z = 0,1)$ denotes the outputs appearing at $SM_z$ side when the atom is in state $|\uparrow\ (\downarrow)\rangle$. $\left|\alpha_{{\rm T},z\uparrow(\downarrow)}^{(m)}\right\rangle$ denotes the transmission field generated by $SSC_z$ in the $m$-th cycle, while $\left|\alpha_{{\rm S},z\uparrow(\downarrow)}^{(m)}\right\rangle$ denotes the scattering field. Given that $|{\rm loss}_{\uparrow(\downarrow)}\rangle$ includes all optical losses caused by the atom-cavity coupling system, the fidelity is calculated by taking the trace of $|{\rm loss}_{\uparrow(\downarrow)}\rangle$, i.e., $F = Tr_{\rm loss}\{\langle\psi_{\rm T}|\psi_{\rm f}\rangle\langle\psi_{\rm f}|\psi_{\rm T}\rangle\} = \{|\langle\alpha|C_{0\uparrow}\rangle|^2 + |\langle-\alpha|C_{0\downarrow}\rangle|^2 + 2{\rm Re}[\langle\alpha|C_{0\uparrow}\rangle\langle C_{0\downarrow}|-\alpha\rangle\langle{\rm loss}_\downarrow|{\rm loss}_\uparrow\rangle]\}/4$. In addition to fidelity $F$, we also investigate cattiness [69-71], $C_{\rm a} = (|\langle{\rm loss}_\uparrow|{\rm loss}_\downarrow\rangle|^2 - |\langle C_{0\uparrow}|C_{0\downarrow}\rangle|^2)|C_{0\uparrow} - C_{0\downarrow}|^2/4$. Detailed calculations can be found in the "Cattiness calculation" subsection in the Methods. This physical quantity characterizes the purity of the prepared CS. Its maximum value is equal to $|\alpha|^2$. Apparently, $C_a$ is more sensitive to light loss compared to $F$, since its value is related to $|\langle{\rm loss}_\downarrow|{\rm loss}_\uparrow\rangle|^2$, while $F$ is solely related to $|\langle{\rm loss}_\downarrow|{\rm loss}_\uparrow\rangle|$.

As a comparison, we simulate the single reflection model [37] as well. More specifically, the input $|\alpha\rangle$ is directly reflected by $\rm SSC_1$, and the corresponding output state is $\left(|\alpha_{{\rm R},1\uparrow}\rangle|{\rm loss}_\uparrow\rangle|\uparrow\rangle + |\alpha_{{\rm R},1\downarrow}\rangle|{\rm loss}_\downarrow\rangle|\downarrow\rangle\right)/\sqrt{2}$ with $|{\rm loss}_{\uparrow(\downarrow)}\rangle = |\alpha_{{\rm T},1\uparrow(\downarrow)}\rangle|\alpha_{{\rm S},1\uparrow(\downarrow)}\rangle$. In this model, the constraints on the atomic parameters $\gamma$ and $\Delta$ can be directly obtained from Eq. (2). For the empty cavity case (atom is in $|\downarrow\rangle$), as long as $\kappa_{\rm T} = 0$, the ideal reflection $\alpha_{\rm R} = -\alpha_{\rm i}$ can be obtained. As for the case where the atom is in $|\uparrow\rangle$, the condition for ideal reflection $\alpha_{\rm R} = \alpha_{\rm i}$ is $\Delta = \gamma = \kappa_{\rm T} = 0$. If only $\gamma$ is non-zero, we can see that the ideal reflection can be approximately achieved only when $\gamma \ll g^2/\kappa_{\rm R}$. As $\gamma$ increases, the cavity reflectivity $|\eta_{{\rm R},1\uparrow}|^2$ decreases monotonically until it drops to 0 when $\gamma = g^2/\kappa_{\rm R}$. In other words, $\gamma$ causes of the first type of light loss. If we focus on $\Delta$, however, it only affects $\beta_{{\rm R},1\uparrow}$ when $\gamma = \kappa_{\rm T} = 0$, since $|\eta_{{\rm R},1\uparrow}|^2 = 1$. As $\Delta$ varies from $-\infty$ to $\infty$, $\beta_{{\rm R},1\uparrow}$ decreases monotonically from $\pi$ to $-\pi$. In order to ensure that $\beta_{{\rm R},1\uparrow}$ is close to 0, the constraint $\Delta \ll g^2/\kappa_{\rm R}$ is required. Our subsequent numerical simulations will demonstrate that satisfying the constraints $\gamma \ll g^2/\kappa_{\rm R}$ and $\Delta \ll g^2/\kappa_{\rm R}$ is necessary for the single reflection model to function properly. In our multiple reflection scheme, however, the above constraints are relaxed.

**Effect of atomic spontaneous emission decay rate $\gamma$**

The atomic parameter $\gamma$ is the primary source of the first type of light loss. To analyze its effect, we plot the fidelity $F$ and cattiness $C_{\rm a}$ against $\tilde{\gamma} = \kappa_{\rm R}\gamma/g^2$ with $g = 2\pi \times 7.8{\rm MHz}$, $\kappa_{\rm R} = \kappa = 2\pi \times 2.3{\rm MHz}$ and $\Delta = 0$ in Fig. 2, where $1/\tilde{\gamma}$ is the cooperativity. In this simulation, we assume that all classical optical devices such as SM are ideal. The yellow dotted-dashed curve is plotted for the single reflection model with $|\alpha|^2 = 4$, which has almost reached the upper limit of such model [37]. The remaining curves illustrate the multiple reflection model. Red curves correspond to $|\alpha|^2 = 4$, black to $|\alpha|^2 = 10$, and blue to $|\alpha|^2 = 16$. Dotted curves are used for $M = 5$, dashed curves for $M = 20$, solid curves for $M = 100$, and double-dotted-dashed curves for $M = 1000$. In addition, $v_{\rm max}$ is plotted in Fig.2c for $M = 20$, where $v_{\rm max} = \max\left\{\left|\alpha_{{\rm i},1\uparrow}^{(1)}\right|^2, \left|\alpha_{{\rm i},1\uparrow}^{(2)}\right|^2, \ldots, \left|\alpha_{{\rm i},1\uparrow}^{(m)}\right|^2 \ldots\right\}$ is the maximum value of the average photon number

reaching $SSC_1$ in each single cycle across all cycles when the atom is in $|\uparrow\rangle$. As shown in the figure, $v_{\max}$ is always less than 1 (For other $M$, the situation is similar), which validates the low atomic excitation probability condition, hence Eqs. (2)-(4) are valid for simulations.

By comparison, we can see that the multiple reflection scheme outperforms the single reflection scheme. In our scheme, it is evident that fidelity increases as $M$ increases. Whereas for larger $|\alpha|^2$, larger $M$ is required to achieve the same fidelity. More importantly, for $\gamma$ much larger than $2\pi \times 3.0\mathrm{MHz}$ (This value is taken from the experiment [37]. It corresponds to $\tilde{\gamma} \approx 0.11$ and has been marked in the figure), our scheme can still provide large $F$. To better explain the result, we consider the extreme case when $\tilde{\gamma} = 1$, which means all photons reaching $\mathrm{SSC}_1$ in a single cycle are lost when the atom is in $|\uparrow\rangle$. Under such conditions, the interference between Zone 0 and Zone 1 is continuously interrupted, resulting in the output light field state in Zone 0 becomes $|\alpha \cos^2(\pi/2M) \cos^{M-1} 2\,(\pi/2M)\rangle \approx |\alpha(1 - \pi^2/2M)\rangle$ [48,49], where the approximation is valid when $M$ is sufficiently large. It is evident that when $M \gg |\alpha|$, the output photon state approaches $|\alpha\rangle$, signifying that the total light loss induced by the atom-cavity coupling system over the entire preparation process ( $M$ cycles) is actually significantly suppressed. This phenomenon is a manifestation of the quantum Zeno effect, which is caused by photon loss in the individual atom-cavity coupling system. We note that the cumulative loss is directly proportional to $|\alpha|^2$ and inversely proportional to $M$, which explains why in Fig.2a, the larger $M$ and the smaller $|\alpha|^2$, the better the fidelity. So far, our discussion is about $\tilde{\gamma} = 1$. As for the case of $0 < \tilde{\gamma} < 1$, the situation is similar. There is a mixture of two physical mechanisms. The first is to maintain the initial state by phase modulation, which does not cause any photon loss. The second is the quantum Zeno effect, which brings photon loss to the entire system. As $\tilde{\gamma}$ increases, the role of the second mechanism becomes increasingly important. Fortunately, by increasing $M$, the total loss can be greatly suppressed, which indicates that the quantum Zeno effect determines the lower bound of fidelity of our scheme. Together, the above two mechanisms ensure that our scheme has higher fidelity and higher tolerance to $\gamma$ than the single reflection scheme as $M$ increases. In addition, since the condition $\gamma \ll g^2/\kappa_{\mathrm{R}}$ is relaxed, it implies that our scheme does not require strong coupling between atom and cavity when the value of $\gamma$ is fixed.

Regarding cattiness, due to its greater sensitivity to light loss, achieving high cattiness requires much smaller light loss than the fidelity case. This necessitates a larger value of $M$. Consequently, we have plotted the blue double-dotted dashed curve for $|\alpha|^2 = 16, M = 1000$ in Fig.2b. It is evident that even when the light loss in an individual atom-cavity coupling system is $100\%$, the cattiness can still exceed 10. In conjunction with our discussion on fidelity, even for significantly large values of $|\alpha|^2$, our scheme can ensure high fidelity and cattiness, provided the condition $M \gg |\alpha|$ is satisfied. This means that Type I loss does not limit the size of CS that our scheme can produce.

**Effect of atom-cavity detuning Δ**

Following the analysis of $\gamma$, we discuss the impact of $\Delta$. We have shown that by interrupting the interference, the transmission of the light field from Zone 0 to Zone 1 can be suppressed. Note that the phase mismatch between the two Zones also interrupts the interference, we expect that our scheme can have high tolerance of $\Delta$ as well. In Fig. 3, we plot fidelity $F$ and cattiness $C_{\mathrm{a}}$ against $\tilde{\Delta} = \kappa_{\mathrm{R}}\Delta/g^2$. Solid curves are plotted for $\gamma = 0$. Dashed curves are for $\gamma = 2\pi \times 3.0\mathrm{MHz}$. The values of $g, \kappa_{\mathrm{R}}$ and $\kappa_{\mathrm{T}}$ are the same as in Fig. 2. In addition, the yellow curves are plotted for the single reflection model with $|\alpha|^2=4$. As for the multiple reflection model, the red curves are for $|\alpha|^2=4, M=5$, the black curves are for $|\alpha|^2=10, M=20$ and the blue curves are for $|\alpha|^2=16, M=100$, respectively. As shown in Fig.3, even for large $|\alpha|$, as long as $M$ is large, our scheme can be less sensitive to Δ. Moreover, since Δ does not induce optical losses, the responses of $F$ and $C_{\mathrm{a}}$ to different $M$ are consistent.

**Effect of the atomic decoherence**

Besides the atomic parameters $\gamma$ and Δ, next we provide a discussion about the influence of the decoherence between the atomic states $|\downarrow\rangle$ and $|\uparrow\rangle$. Obviously, our scheme requires the atom to remain in superposition at least until the end of $M$ cycles. Nevertheless, we need to mention that the multiple reflection processes hardly affect the atomic decoherence. In our scheme, when the atom is in $|\uparrow\rangle$, the low atomic excitation probability can be satisfied. As for the atom in $|\downarrow\rangle$, it is not coupled to the light field. We note that the atomic superposition state has been reported to last about 400μs [75,76], whereas the full-width at half-maximum of the light pulse that is employed in the experiment of single reflection model is 2.3μs [37]. Therefore, it is possible for our scheme to be completed before the decoherence.

**Effect of light loss caused by classical optical devices $\varepsilon$**

So far, we have primarily discussed atomic parameters and optical losses due to atom (Type I loss). Next, we explore the impact of Type II loss, caused by classical optical devices, while keeping Type I light loss constant. We postulate that the loss rate per cycle outside the cavity is denoted by $\varepsilon$. Accordingly, the cumulative loss rate over the entire preparation process can be approximated as $M\varepsilon$, given that $(1-\varepsilon)^M \approx 1-M\varepsilon$ under the assumption of a small $\varepsilon$. Furthermore, considering the symmetrical of the optical devices on both arms of the interferometer, it is reasonable to assume that Type II loss is identical for each arm. Consequently, $\varepsilon$ can be expressed as the sum of the light losses from each optical device located on a single arm of the interferometer. To facilitate our discussion, we assume that all optical devices are ideal, except for the switchable mirror SM (We also assume that irrespective of whether SM is turned on or off, its light loss rate remains constant). When coherent light $|\beta_z\rangle$ incident upon $\mathrm{SM}_z$ ($z=0,1$), the reflected field is denoted as $|\sqrt{1-\varepsilon}\beta_z\rangle$, and the lost light field can be represented as $|\sqrt{\varepsilon}\beta_z\rangle$. In our simulation, the effect of these loss terms $|\sqrt{\varepsilon}\beta_z\rangle$ from different cycles is accounted for by taking the trace, similar to our treatment of the cavity mentioned earlier.

In Fig.4, we plot $\varepsilon$ against effective fidelity $F_{\mathrm{ef}}=Tr_{\mathrm{loss}}\{\langle\psi_{\mathrm{ef}}|\psi_{\mathrm{f}}\rangle\}$, cattiness $C_{\mathrm{a}}$, and output intensity $|\alpha_{\mathrm{ef}}|^2$ for different $|\alpha|^2$and $M$. Here, $F_{\mathrm{ef}}$ differs from $F$ in that the target state is set as $|\psi_{\mathrm{ef}}\rangle=(|\alpha_{\mathrm{ef}}\rangle|\uparrow\rangle+|-\alpha_{\mathrm{ef}}\rangle|\downarrow\rangle)/\sqrt{2}$ with $\alpha_{\mathrm{ef}}=-C_{0\downarrow}=\alpha\,[(\kappa_{\mathrm{R}}-\kappa_{\mathrm{T}})/(\kappa_{\mathrm{T}}+\kappa_{\mathrm{R}})]^M$. Note that $|C_{0\downarrow}\rangle$ is the output when the atom is in $|\downarrow\rangle$, where interference occurs between the two empty cavities. If the optical parameters of these two cavities are the same, only intensity of the output is affected and reduced from $|\alpha|^2$ to $|\alpha_{\mathrm{ef}}|^2$. For the cavity-atom system, the parameters are set as $g=2\pi\times7.8\mathrm{MHz}$, $\kappa_{\mathrm{R}}=2\pi\times2.3\mathrm{MHz}$, and $\gamma=2\pi\times3.0\mathrm{MHz}$, aligning with actual cavity characteristics. However, for $\kappa_{\mathrm{T}}$, we set it to zero. Based on the above parameters, the corresponding Type I loss is about 36.6%. Before we begin analyzing the impact of Type II light loss, it is important to note that the role of $\kappa_{\mathrm{T}}$ can be simulated with $\varepsilon$, since the main impact of $\kappa_{\mathrm{T}}$ is on the empty cavity light loss. This loss can be added to other Type II losses. Therefore, here we simply treat it as 0. Detailed explanations regarding $\kappa_{\mathrm{T}}$ are provided later.

For comparison, we have also plotted the curves corresponding to the single reflection model with $|\alpha|^2 = 4$ (the yellow dotted-dashed curves). In this scenario, both $F_{\mathrm{ef}}$ and $C_{\mathrm{a}}$ demonstrate minimal impact from $\varepsilon$, indicating that the key factor affecting the single reflection model here primarily stems from the optical losses in the individual atom-cavity system (Type I light loss). In this context, we utilize the yellow curve as a baseline to assess the quality of CS preparation in the multiple reflection model. From Fig.4a and 4b, it can be observed that when $\varepsilon$ is large, the multiple reflection model suffers from Type II light loss, with both $F_{\mathrm{ef}}$ and $C_{\mathrm{a}}$ falling below that of the single reflection model. As $M$ increases, the situation deteriorates further. However, as $\varepsilon$ decreases, a dramatic shift occurs. Both $F_{\mathrm{ef}}$ and $C_{\mathrm{a}}$ begin to outperform the single reflection case and rapidly increase. Smaller values of $|\alpha|^2$ and $M$ exhibit a higher tolerance to Type II light loss, but larger $M$ values can achieve greater maximal values of $F_{\mathrm{ef}}$ and $C_{\mathrm{a}}$ (at $\varepsilon = 0$). As a result, there are observable intersections of curves for different $M$ values for a given $|\alpha|^2$, which implies that it's possible to obtain better fidelity using higher $M$. For larger values of $|\alpha|^2$, such improvements due to $M$ are only noticeable when $\varepsilon$ is closer to zero. All these observed effects are because that the impact of Type I losses gradually becomes dominant as $\varepsilon$ approaches zero, and our method effectively suppresses the influence of Type I light loss. It's important to recognize that in the single reflection model, Type I light losses can only be reduced through improvements in the atom-cavity system. In contrast, our multiple reflection approach, while not immune to light losses from classical optical devices, offers a method to reduce the stringent requirements on quantum microscopic system. This is the central and distinctive feature of our work.

To pave the way for the initiation of proof-of-concept experiments, we also provide two tables specifying the loss tolerances for classical optical devices in a single cycle. These loss tolerances are required to achieve certain levels of effective fidelity and cattiness under varying conditions of $|\alpha|^2$ and $M$. The tables reveal that, to achieve over 70% effective fidelity in conditions with $|\alpha|^2 \leq 4$ and $M \leq 10$, our scheme's tolerance for Type II loss can be close to or even exceed 2%. For larger $|\alpha|^2$, however, our scheme requires much smaller $\varepsilon$. Hence, for our scheme, the constraint on preparing large-sized CS is tied to the quality of classical optical devices.

**Discussion on the cavity parameter $\kappa_{\mathrm{T}}$ and potential approaches to reduce light loss in empty cavity**

Next, we provide additional explanation and clarification on the role of $\kappa_{\mathrm{T}}$. We note that $\kappa_{\mathrm{T}}$ causes light loss when the atom is in either $|\uparrow\rangle$ or $|\downarrow\rangle$, thus affecting both the first and second types of light loss. However, when the atom is in $|\uparrow\rangle$, according to Eqs. (3)-(4), and the fact that $g > \gamma \gg \kappa_{\mathrm{T}}$ [37], the loss caused by light field passing through the atom-cavity coupling system (Eq. (3)) is much smaller than the light loss caused by atomic scattering (Eq. (4)). As for Eq. (4), since $\kappa_{\mathrm{T}} \ll \kappa_{\mathrm{R}}$, the influence of $\kappa_{\mathrm{T}}$ can be neglected. Hence, the Type I light loss due to $\kappa_{\mathrm{T}}$ is not our primary concern. In the scenario where the atom is in $|\downarrow\rangle$ and does not participate in the interaction, the light transmission through the empty cavity falls under Type II light loss. In other words, the effect of $\kappa_{\mathrm{T}}$ can be accounted for in $\varepsilon$. According to Eq. (2), the light loss rate of the empty cavity after a single reflection is $1 - |(\kappa_{\mathrm{T}} - \kappa_{\mathrm{R}})/(\kappa_{\mathrm{T}} + \kappa_{\mathrm{R}})|^2$. In the experiment [37], the cavity parameters are $\kappa_{\mathrm{T}} = 2\pi \times 0.2$ MHz and $\kappa_{\mathrm{R}} = 2\pi \times 2.3$ MHz, which results in $\varepsilon \approx 29\%$ even if other optical devices are ideal, while after a few reflections, almost all photons are lost. Therefore, this cavity is unfortunately not suitable for our scheme. To reduce the optical losses, one needs either decrease $\kappa_{\mathrm{T}}$ or increase $\kappa_{\mathrm{R}}$. The latter is simpler in practice. However, although increasing $\kappa_{\mathrm{R}}$ can reduce photon loss during the interference of two empty cavities (The atom is in state $|\downarrow\rangle$), it also increases photon loss in the presence of atom-cavity coupling (The atom is in state $|\uparrow\rangle$). To verify this, we plot Fig.5 with $|\alpha|^2 = 8$, $\gamma = 2\pi \times 3.0\mathrm{MHz}$, and $\Delta = 0$. In Fig.5a (for $M = 10$) and 5b (for $M = 50$), we plot the effective fidelity $F_{\mathrm{ef}}$ against $\kappa_{\mathrm{R}}$. Solid curves represent $\kappa_{\mathrm{T}} = 2\pi \times 0.02\mathrm{MHz}$, while dashed curves represent $\kappa_{\mathrm{T}} = 2\pi \times 0.002\mathrm{MHz}$. The corresponding light loss rate of the empty cavity is approximately 3.4% for $\kappa_{\mathrm{T}} = 2\pi \times 0.02\mathrm{MHz}$ and 0.35% for $\kappa_{\mathrm{T}} = 2\pi \times 0.002\mathrm{MHz}$ when $\kappa_{\mathrm{R}} = 2\pi \times 2.3\mathrm{MHz}$. Black curves are for $g = 2\pi \times 7.8\mathrm{MHz}$, red for $g = 2\pi \times 15\mathrm{MHz}$, and blue for $g = 2\pi \times 30\mathrm{MHz}$. As for Fig.5c and 5d, the green curves are plotted for $|\alpha_{\mathrm{ef}}|^2$, while curves of other colors represent $C_{\mathrm{a}}$. It is observed that $F_{\mathrm{ef}}$ can be significantly improved as $\kappa_{\mathrm{T}}$ decreases. Regarding $\kappa_{\mathrm{R}}$, as it initially increases, $|\alpha_{\mathrm{ef}}|^2$ rapidly rises to its maximum value 8, which causes $F_{\mathrm{ef}}$ to increase. Subsequently, photon loss due to atom-cavity coupling plays a major role, resulting in the decrease of $F_{\mathrm{ef}}$. Particularly, it's noteworthy that for the black solid curve, when $F_{\mathrm{ef}}$ starts to decrease, its corresponding $|\alpha_{\mathrm{ef}}|^2$ is not close to 8. The reason is that $\kappa_{\mathrm{R}}$ is approaching the limit $g^2/\gamma$. Under such limit, the photon loss of a single reflection on $\mathrm{SSC}_1$ when the atom is in $|\uparrow\rangle$ is almost 100%. Therefore, we choose larger values of $g$ to raise the limit so that the corresponding $|\alpha_{\mathrm{ef}}|^2$ of maximum $F_{\mathrm{ef}}$ can get closer to the maximum value 8, thereby increasing $F_{\mathrm{ef}}$. This phenomenon is verified in Fig.5. Moreover, in the blue curves, $\kappa_{\mathrm{R}}$ maintains a wide range of high fidelity. This is due to relaxed constraint of $g^2 \gg \kappa_{\mathrm{R}}\gamma$ in our scheme. However, we must emphasize that achieving high fidelity does not necessarily require increasing $g$. By decreasing $\kappa_{\mathrm{T}}$, we can achieve the same purpose. In fact, the motivation of this work is to reduce the influence of the atom, and to show that the performance of our protocol can be improved by just upgrading the classical optical system, such as the parameters $\kappa_{\mathrm{T}}$ and $\varepsilon$.

Regarding the impact of $M$, we provide a brief explanation here. When $\kappa_{\mathrm{R}}$ is small, significant Type II light loss leads to a decrease in our scheme's performance with larger $M$ values, as depicted in Fig. 5. However, as $\kappa_{\mathrm{R}}$ increases, leading to a decrease in Type II light loss, larger $M$ values can ultimately offer improved $F_{\mathrm{ef}}$ and $C_{\mathrm{a}}$.

**Discussion on pulse length and interferometer size**

Based on the above analysis of various experimental parameters, we now discuss the requirements of our scheme for the duration of the incident pulse. It is noteworthy that the influence of the atom-cavity coupling system on the wave packet waveshape during a single reflection process has been discussed [42]. The results indicate that to shorten the pulse length by half, while maintaining the reflected waveshape unchanged is necessary to double $\kappa_{\mathrm{R}}$. For the single reflection model, an increase in $\kappa_{\mathrm{R}}$ amplifies Type I losses, resulting in reduced fidelity, hence is not a favorable option. In our scheme, however, since Type I losses can be mitigated, increasing $\kappa_{\mathrm{R}}$ is feasible. For example, in the experiment [37], the pulse duration is $T_{\mathrm{p}} = 2.3\mu\mathrm{s}$, and the cavity parameter $\kappa_{\mathrm{R}}$ is $2\pi \times 2.3\mathrm{MHz}$. According to our simulations in Fig. 5, when $|\alpha|^2 = 8$, $M = 50$, $\kappa_{\mathrm{T}} = 2\pi \times 0.002\mathrm{MHz}$, and $\kappa_{\mathrm{R}} = 2\pi \times 10\mathrm{MHz}$, the effective fidelity of the output CS in our scheme is about $0.75$. With $\kappa_{\mathrm{R}}$ increased by approximately 5 times, the pulse length used in our scheme can be shortened 5 times under the condition of keeping waveshape unchanged, i.e., to $T_{\mathrm{p}} \approx 0.5\mu\mathrm{s}$.

We, so far, stipulated that the waveshape of the reflected pulse remains unchanged. This is a requirement from the single reflection model, but not a necessity in our scheme. This is due to the fact that when the atom is in $|\downarrow\rangle$, interference occurs between the two empty cavities, and even if the reflected waveshape distorts, it does not affect the interference process. Conversely, when the atom is in $|\uparrow\rangle$, the role of the atom is merely to disrupt the interference. In this case, the light field is primarily concentrated on the empty cavity side, and the waveshape of the reflected pulse is still determined by the empty cavity reflection. Consequently, variations in the waveshape of the output pulse do not affect the generation of CS. This facilitates further compression of the pulse length in our scheme.

Next, we briefly discuss the size of the interferometer. If we define the size of the interferometer by the distance $L$ between $\mathrm{SM}_0$ and $\mathrm{SSC}_0$, to ensure that the light-matter interaction in two cycles does not interfere with each other, it implies that $L$ needs to be greater than the half pulse length, i.e., $L \geq (T_\mathrm{p}c)/2$, where $c$ is the speed of light. Hence, the total flight distance of the pulse in the entire CS preparation process is $MT_\mathrm{p}c$. For a pulse duration $T_\mathrm{p} = 2.3\mathrm{\mu s}$, $L$ needs to be at least $350\mathrm{m}$, and the total flight distance falls within the kilometer scale. Given such distances, employing optical fibers to construct the interferometer is a reasonable choice. Regarding the requirements for light loss from classical optical devices, one can refer to Tables 1 and 2 we provided. We note that the parameter $\kappa_\mathrm{R}$ in these tables is specified as $2\pi \times 2.3\mathrm{MHz}$. Increasing it allows us to further compress the pulse duration. For instance, setting $\kappa_R$ to $2\pi \times 10\mathrm{MHz}$ results in a pulse duration $T_\mathrm{p}$ of $0.5\mathrm{\mu s}$, with minimum $L$ adjusted to $75\mathrm{m}$. By fixing other parameters at $(g, \gamma) = 2\pi \times (7.8, 3.0)\mathrm{MHz}$ and $\kappa_\mathrm{T} = \Delta = 0$, along with using $M = 20$ and the best-reported optical fiber, which has an average attenuation of less than $0.157\ \mathrm{dB\ km^{-1}}$ [77] (the corresponding Type II loss is about $0.53\%$), the effective fidelity of the output CS is about $0.77$ for an input light field of $|\alpha|^2 = 3$. In this estimate, we have not yet accounted for other classical optical losses, such as those from SM and fiber-cavity coupling. If we set these additional losses to $2\%$, the effective fidelity reduces to $0.63$. While our scheme can shift the source of errors from the atomic system to classical optical systems, it also demands higher manufacturing technology standards for existing optical devices.

Lastly, we address the requirements for the SM. In addition to light loss restrictions, there are also requirements for its response time $T_\mathrm{s}$. One promising candidate of SM is Pockels cell [65-68]. Applying high voltage, it can rotate the polarization of the light field by 90 degrees, facilitating the extraction of the light field from the interferometer. Its response time is in the nanosecond range. Hence, $T_\mathrm{s}$ can be significantly shorter than the duration of the pulse $T_\mathrm{p}$. Strictly speaking, to ensure that the transmission of the pulse is uninterrupted, the length of $L$ should be $(T_\mathrm{p} + T_\mathrm{s})c/2$, which means there needs to be a sufficient time window allowed for the change of state in the Pockels Cell. However, since $T_\mathrm{s} \ll T_\mathrm{p}$, $L$ is primarily determined by $T_\mathrm{p}$.

## III. Conclusions

We propose a method for generating macro-micro entanglement based on the interaction-free measurement and the quantum Zeno effect. Our method allows for the deterministic preparation of optical CS with arbitrarily large average photon numbers. Because only a small fraction of photons actually interacts with the quantum microsystem during multiple interactions, our method avoids directly exposing the quantum microsystem to a strong light field. This indirect control mechanism is a key feature of our work. Another significant feature of our work is the tolerance it shows for photon loss within the quantum microsystem. We show that the state $|-\alpha\rangle$ is generated through multiple optical interferences, independent of the quantum microsystem.

Conversely, the state $|\alpha\rangle$ results from the disruption of the interference process, which remains feasible even if the quantum microsystem experiences substantial photon loss or other imperfections. This fact enables our scheme to improve the fidelity of the output CS through enhancements to the classical optical system.

## IV. Methods

### A. System wave-function calculation

We have mentioned that $a_0$ and $a_1$ represent the annihilation operators of the light field in Zone 0 and Zone 1, respectively. Based on this, the function of $BS$ can be described as $a_0^\dagger \rightarrow a_0^\dagger \cos\theta + a_1^\dagger \sin\theta$ and $a_1^\dagger \rightarrow a_1^\dagger \cos\theta - a_0^\dagger \sin\theta$ where $\theta = \pi/2M$ . Now, we consider an arbitrary initial photon state

$$|\text{Initial}\rangle = |u, v\rangle = \exp\left(ua_0^\dagger - u^* a_0\right)\exp\left(va_1^\dagger - v^* a_1\right)|0,0\rangle, \quad (5)$$

which represents that a coherent state $|u\rangle$ is in Zone 0 and a coherent state $|v\rangle$ is in Zone 1. After passing through the BS, we have the final state

$$\begin{aligned}|\text{Final}\rangle &= \exp\left[u\left(a_0^\dagger \cos\theta + a_1^\dagger \sin\theta\right) - u^*(a_0 \cos\theta + a_1 \sin\theta)\right], \\ &\quad \times \exp\left[v\left(a_1^\dagger \cos\theta - a_0^\dagger \sin\theta\right) - v^*(a_1 \cos\theta - a_0 \sin\theta)\right]|0,0\rangle \\ &= |u\cos\theta - v\sin\theta, u\sin\theta + v\cos\theta\rangle. \end{aligned} \quad (6)$$

Similarly, consider an arbitrary phase operation $a^\dagger \rightarrow e^{i\varphi}a^\dagger$. For the initial state $|\text{Initial}\rangle = |u\rangle$, after the operation, the final state is

$$\begin{aligned}|\text{Final}\rangle &= \exp\left(-\frac{1}{2}|u|^2\right)\sum_{n=0}^{\infty}\frac{u^n}{\sqrt{n!}}\frac{1}{\sqrt{n!}}\left(e^{i\varphi}a^\dagger\right)^n|0\rangle \\ &= \exp\left(-\frac{1}{2}|u|^2\right)\sum_{n=0}^{\infty}\frac{1}{\sqrt{n!}}\left(ue^{i\varphi}\right)^n|n\rangle = |ue^{i\varphi}\rangle \end{aligned} \quad (7)$$

Based on Eqs. (6) and (7), we provide the calculation of Eq. (1).

At the beginning of the preparation, the wave-function of the whole system is

$$\left|\psi^{(0)}\right\rangle = \frac{\sqrt{2}}{2}|\alpha, 0\rangle(|\uparrow\rangle + |\downarrow\rangle) \quad (8)$$

In the first cycle, after the photons pass through $BS$ for the first time, the system state is

$$\left|\psi^{(1)}\right\rangle = \frac{\sqrt{2}}{2}|\alpha\cos\theta, \alpha\sin\theta\rangle(|\uparrow\rangle + |\downarrow\rangle) \quad (9)$$

Before the photons are reflected by SSC, the system state becomes $\frac{\sqrt{2}}{2}|-\alpha\cos\theta, -\alpha\sin\theta\rangle(|\uparrow\rangle + |\downarrow\rangle)$ due to PS. Regarding the reflection, we emphasize that only when the atom is in $|\uparrow\rangle$, $SSC_1$ does not change the phase of the reflected field. As a result, the wave function of the whole system becomes $\frac{\sqrt{2}}{2}|\alpha\cos\theta, -\alpha\sin\theta\rangle|\uparrow\rangle + \frac{\sqrt{2}}{2}|\alpha\cos\theta, \alpha\sin\theta\rangle|\downarrow\rangle$ . Subsequently, after the second time that the photons pass through BS, we have

$$\left|\psi^{(2)}\right\rangle = \frac{\sqrt{2}}{2}|\alpha, 0\rangle|\uparrow\rangle + \frac{\sqrt{2}}{2}|\alpha\cos 2\theta, \alpha\sin 2\theta\rangle|\downarrow\rangle \quad (10)$$

This state becomes the initial state of the second cycle, and the process is repeated. It is not difficult to obtain that after $m$ cycles, the wave-function of the whole system is

$$\left|\psi^{(2m)}\right\rangle = \frac{\sqrt{2}}{2}|\alpha, 0\rangle|\uparrow\rangle + \frac{\sqrt{2}}{2}|\alpha\cos 2m\theta, \alpha\sin 2m\theta\rangle|\downarrow\rangle \quad (11)$$

Here the superscript of $\left|\psi^{(2m)}\right\rangle$ represents the photons pass through BS $2m$ times.

### B. Input-output theory for the atom-cavity coupling system

The Hamiltonian of cavity-atom system ($SSC_1$) can be described as

$$H = \hbar\omega_e\sigma_{ee} + \hbar\omega_\uparrow\sigma_{\uparrow\uparrow} + \hbar\omega_c a^\dagger a + \hbar\sum_{J=R,T,S}\int_{-\infty}^{\infty}\omega_J b_J^\dagger(\omega_J)b_J(\omega_J)d\omega_J$$

$$+\hbar g\left(\sigma_{\uparrow e}a^\dagger + \sigma_{e\uparrow}a\right) + \hbar\sqrt{\frac{\gamma}{\pi}}\int_{-\infty}^{\infty}\left[\sigma_{\uparrow e}b_S^\dagger(\omega_S) + \sigma_{e\uparrow}b_S(\omega_S)\right]d\omega_S$$

$$+i\hbar\sqrt{\frac{\kappa_R}{\pi}}\int_{-\infty}^{\infty}\left[ab_R^\dagger(\omega_R) - a^\dagger b_R(\omega_R)\right]d\omega_R + i\hbar\sqrt{\frac{\kappa_T}{\pi}}\int_{-\infty}^{\infty}\left[ab_T^\dagger(\omega_T) - a^\dagger b_T(\omega_T)\right]d\omega_T \quad (12)$$

where $\hbar\omega_e$ is the energy of excited atomic state $|e\rangle$, $\hbar\omega_\uparrow$ is the energy of the atomic state $|\uparrow\rangle$, $\omega_c$ is the frequency of the cavity mode described by annihilation operator $a$, $\omega_J$ is the frequency of external field described by annihilation operator $b(\omega_{J=R,T,S})$ with $\left[b_J(\omega_J), b_J^\dagger(\omega_J^{'})\right] = \delta(\omega_J - \omega_J^{'})$, and the subscript R represents the external multi-mode field on $CM_R$ side, T represents the external field on $CM_T$ side, S represents the scattering field due to the atomic spontaneous emission. In addition, $g$ is coupling constant between the cavity and the atomic transition between $|e\rangle$ and $|\uparrow\rangle$, $2\gamma$ is the spontaneous atomic decay rate on the same transition, $\kappa_R$ and $\kappa_T$ are cavity field decay rates. We also set that $\sigma_{\uparrow e} = |\uparrow\rangle\langle e|$, $\sigma_{ee} = |e\rangle\langle e|$ and $\sigma_{\uparrow\uparrow} = |\uparrow\rangle\langle\uparrow|$.

Based on the above Hamiltonian, it is not difficult to obtain the following Heisenberg equations

$$\frac{da(t)}{dt} = -i\omega_c a(t) - ig\sigma_{\uparrow e}(t) - \sum_{J=R,T}\sqrt{\frac{\kappa_J}{\pi}}\int_{-\infty}^{\infty}b_J(\omega_J,t)d\omega_J, \quad (13)$$

$$\frac{d}{dt}\sigma_{\uparrow e}(t) = -i(\omega_e - \omega_\uparrow)\sigma_{\uparrow e}(t) + ig[\sigma_{ee}(t) - \sigma_{\uparrow\uparrow}(t)]a(t)$$

$$+i\sqrt{\frac{\gamma}{\pi}}\int_{-\infty}^{\infty}[\sigma_{ee}(t) - \sigma_{\uparrow\uparrow}(t)]\, b_S(\omega_S,t)d\omega_S$$

$$\approx -i(\omega_e - \omega_\uparrow)\sigma_{\uparrow e}(t) - iga(t) - i\sqrt{\frac{\gamma}{\pi}}\int_{-\infty}^{\infty}b_S(\omega_S,t)d\omega_S. \quad (14)$$

In the approximation, we have assumed that [37,45,72]

$$\langle(\sigma_{ee} - \sigma_{\uparrow\uparrow})a\rangle = -\langle a\rangle, \quad (15)$$

which indicates that the atom stays in the state $|\uparrow\rangle$ most of the time. This can be satisfied when the input is weak.

In addition, we can also obtain Heisenberg equations for $b(\omega)$. They are

$$\frac{db_J(\omega_J,t)}{dt} = -i\omega_J b_J\left(\omega_J,t\right) + \sqrt{\frac{\kappa_R}{\pi}}a(t), \quad J = R, T, \quad (16)$$

$$\frac{db_S(\omega_S,t)}{dt} = -i\omega_S b_S(\omega_S,t) - i\sqrt{\frac{\gamma}{\pi}}\sigma_{\uparrow e}(t). \quad (17)$$

Eqs. (16) and (17) can be rewritten in integral form. If we assume that the atom-light interaction begins at time $T_{in} < t$, we have

$$b_J(\omega_J,t) = b_J(\omega_J,T_{in})e^{i\omega_J(T_{in}-t)} + \sqrt{\frac{\kappa_J}{\pi}}\int_{T_{in}}^{t}a(t')e^{i\omega_J(t'-t)}dt', \quad (18)$$

$$b_S(\omega_S,t) = b_S(\omega_S,T_{in})e^{i\omega_S(T_{in}-t)} - i\sqrt{\frac{\gamma}{\pi}}\int_{T_{in}}^{t}\sigma_{\uparrow e}(t')e^{i\omega_S(t'-t)}dt'. \quad (19)$$

If we assume that the atom-light interaction ends at time $T_{out} > t$, we have

$$b_J(\omega_J,t) = b_J(\omega_J,T_{out})e^{i\omega_J(T_{out}-t)} - \sqrt{\frac{\kappa_J}{\pi}}\int_{t}^{T_{out}}a(t')e^{i\omega_J(t'-t)}dt', \quad (20)$$

$$b_S(\omega_S,t) = b_S(\omega_S,T_{out})e^{i\omega_S(T_{out}-t)} + i\sqrt{\frac{\gamma}{\pi}}\int_{t}^{T_{out}}\sigma_{\uparrow e}(t')e^{i\omega_S(t'-t)}dt'. \quad (21)$$

By integrating Eq. (18) with frequency, it is not difficult to obtain that

$$\sqrt{\frac{\kappa_J}{\pi}}\int_{-\infty}^{\infty} b_J(\omega_J,t)d\omega_J$$

$$=\sqrt{\frac{\kappa_J}{\pi}}\int_{-\infty}^{\infty} b_J(\omega_J,T_{\mathrm{in}})\mathrm{e}^{\mathrm{i}\omega_J(T_{\mathrm{in}}-t)}d\omega_J+2\kappa_J\int_{T_{\mathrm{in}}}^{t} a(t')dt'\frac{1}{2\pi}\int_{-\infty}^{\infty}\mathrm{e}^{\mathrm{i}\omega_J(t'-t)}d\omega_J$$

$$=\sqrt{2\kappa_J}a_{J,\mathrm{in}}(t)+\kappa_J a(t). \quad (22)$$

where we have used the relation [74]

$$\int_{t_0}^{t} f(t')\delta(t-t')dt'=\int_{t}^{t_1} f(t')\delta(t-t')dt'=\frac{1}{2}f(t),\quad (t_0<t<t_1), \quad (23)$$

and the assumptions ($J=\mathrm{R},\mathrm{T}$)

$$a_{J,\mathrm{in}}(t)=\frac{1}{\sqrt{2\pi}}\int_{-\infty}^{\infty} b_J(\omega_J,T_{\mathrm{in}})\mathrm{e}^{\mathrm{i}\omega_J(T_{\mathrm{in}}-t)}d\omega_J, \quad (24)$$

$$a_{J,out}(t)=\frac{1}{\sqrt{2\pi}}\int_{-\infty}^{\infty} b_J(\omega_J,T_{\mathrm{out}})\mathrm{e}^{\mathrm{i}\omega_J(T_{\mathrm{out}}-t)}d\omega_J. \quad (25)$$

$$a_{\mathrm{S,in}}(t)=\frac{1}{\sqrt{2\pi}}\int_{-\infty}^{\infty} b_{\mathrm{S}}(\omega_{\mathrm{S}},T_{\mathrm{in}})\mathrm{e}^{\mathrm{i}\omega_{\mathrm{S}}(T_{\mathrm{in}}-t)}d\omega_{\mathrm{S}}, \quad (26)$$

$$a_{\mathrm{S,out}}(t)=\frac{1}{\sqrt{2\pi}}\int_{-\infty}^{\infty} b_{\mathrm{S}}(\omega_{\mathrm{S}},T_{\mathrm{out}})\mathrm{e}^{\mathrm{i}\omega_{\mathrm{S}}(T_{\mathrm{out}}-t)}d\omega_{\mathrm{S}}. \quad (27)$$

Similarly, from Eqs. (19)-(21), we have

$$\sqrt{\frac{\gamma}{\pi}}\int_{-\infty}^{\infty} b_{\mathrm{S}}(\omega_{\mathrm{S}},t)d\omega_{\mathrm{S}}=\sqrt{2\gamma}a_{\mathrm{S,in}}(t)-\mathrm{i}\gamma\sigma_{\uparrow\mathrm{e}}(t), \quad (28)$$

$$\sqrt{\frac{\kappa_J}{\pi}}\int_{-\infty}^{\infty} b_J(\omega_J,t)d\omega_J=\sqrt{2\kappa_J}a_{J,\mathrm{out}}(t)-\kappa_J a(t), \quad (29)$$

$$\sqrt{\frac{\gamma}{\pi}}\int_{-\infty}^{\infty} b_{\mathrm{S}}(\omega_{\mathrm{S}},t)d\omega_{\mathrm{S}}=\sqrt{2\gamma}a_{\mathrm{S,out}}(t)+\mathrm{i}\gamma\sigma_{\uparrow\mathrm{e}}(t). \quad (30)$$

Then, by substituting Eqs. (22)(29) into (13), we can obtain the dynamic equations

$$\frac{da(t)}{dt}=-\mathrm{i}\omega_{\mathrm{c}}a(t)-\mathrm{i}g\sigma_{\uparrow\mathrm{e}}(t)-\sqrt{2\kappa_{\mathrm{R}}}a_{\mathrm{R,in}}(t)-\kappa_{\mathrm{R}}a(t)-\sqrt{2\kappa_{\mathrm{T}}}a_{\mathrm{T,in}}(t)-\kappa_{\mathrm{T}}a(t), \quad (31)$$

$$\frac{da(t)}{dt}=-\mathrm{i}\omega_{\mathrm{c}}a(t)-\mathrm{i}g\sigma_{\uparrow\mathrm{e}}(t)-\sqrt{2\kappa_{\mathrm{R}}}a_{\mathrm{R,out}}(t)+\kappa_{\mathrm{R}}a(t)-\sqrt{2\kappa_{\mathrm{T}}}a_{\mathrm{T,in}}(t)-\kappa_{\mathrm{T}}a(t), \quad (32)$$

$$\frac{da(t)}{dt}=-\mathrm{i}\omega_{\mathrm{c}}a(t)-\mathrm{i}g\sigma_{\uparrow\mathrm{e}}(t)-\sqrt{2\kappa_{\mathrm{R}}}a_{\mathrm{R,in}}(t)-\kappa_{\mathrm{R}}a(t)-\sqrt{2\kappa_{\mathrm{T}}}a_{\mathrm{T,out}}(t)+\kappa_{\mathrm{T}}a(t). \quad (33)$$

By substituting Eqs. (28)(30) into (14), we have

$$\frac{d}{dt}\sigma_{\uparrow\mathrm{e}}(t)=-\mathrm{i}(\omega_{\mathrm{e}}-\omega_{\uparrow})\sigma_{\uparrow\mathrm{e}}(t)-\mathrm{i}ga(t)-\mathrm{i}\sqrt{2\gamma}a_{\mathrm{S,in}}(t)-\gamma\sigma_{\uparrow\mathrm{e}}(t), \quad (34)$$

$$\frac{d}{dt}\sigma_{\uparrow\mathrm{e}}(t)=-\mathrm{i}(\omega_{\mathrm{e}}-\omega_{\uparrow})\sigma_{\uparrow\mathrm{e}}(t)-\mathrm{i}ga(t)-\mathrm{i}\sqrt{2\gamma}a_{\mathrm{S,out}}(t)+\gamma\sigma_{\uparrow\mathrm{e}}(t). \quad (35)$$

In the following, we assume that only the input on $\mathrm{CM_R}$ side is none-zero, i.e., $a_{\mathrm{T,in}}(t)=a_{\mathrm{S,in}}(t)=0$. Then, by subtracting (31) and (32), we can get the relation between the input $a_{\mathrm{R,in}}(t)$ and output $a_{\mathrm{R,out}}(t)$,

$$a_{\mathrm{R,in}}(t)+\sqrt{2\kappa_{\mathrm{R}}}a(t)=a_{\mathrm{R,out}}(t). \quad (36)$$

By subtracting (31) and (33), we have

$$a_{\mathrm{T,out}}(t)=\sqrt{2\kappa_{\mathrm{T}}}a(t). \quad (37)$$

By subtracting (34) and (35), we have

$$a_{\mathrm{S,out}}(t)=-\mathrm{i}\sqrt{2\gamma}\sigma_{\uparrow\mathrm{e}}(t). \quad (38)$$

In addition to the above relations, we next calculate the steady-state solution of the dynamic equations (31)(32)(34) by assuming that the cavity-atom system is driven by the input light field with frequency $\omega$. We suppose that

$$a_{\mathrm{R,in}}(t)=\alpha_{\mathrm{i},1\uparrow}\mathrm{e}^{-\mathrm{i}\omega t},$$

$$a(t)=\alpha\mathrm{e}^{-\mathrm{i}\omega t},$$

$$\sigma_{\uparrow e}(t) = \tilde{\sigma} e^{-i\omega t},$$
$$a_{R,out}(t) = \alpha_{R,1\uparrow} e^{-i\omega t},$$
$$a_{T,out}(t) = \alpha_{T,1\uparrow} e^{-i\omega},$$
$$a_{S,out}(t) = \alpha_{S,1\uparrow} e^{-i\omega t}. \tag{39}$$

Then, we can obtain that

$$[-i(\omega_c - \omega) - \kappa_R - \kappa_T]\alpha - ig\tilde{\sigma} - \sqrt{2\kappa_R}\alpha_{i,1\uparrow} = 0, \tag{40}$$
$$[-i(\omega_c - \omega) + \kappa_R - \kappa_T]\alpha - ig\tilde{\sigma} - \sqrt{2\kappa_R}\alpha_{R,1\uparrow} = 0, \tag{41}$$
$$[-i(\omega_e - \omega_\uparrow - \omega) - \gamma]\tilde{\sigma} - ig\alpha = 0. \tag{42}$$

It is not difficult to get that

$$\alpha_{i,1\uparrow} = -\frac{[i(\omega_c-\omega)+\kappa_R+\kappa_T][i(\omega_e-\omega_\uparrow-\omega)+\gamma]+g^2}{\sqrt{2\kappa_R}[i(\omega_e-\omega_\uparrow-\omega)+\gamma]}\alpha, \tag{43}$$
$$\alpha_{R,1\uparrow} = -\frac{[i(\omega_c-\omega)-\kappa_R+\kappa_T][i(\omega_e-\omega_\uparrow-\omega)+\gamma]+g^2}{\sqrt{2\kappa_R}[i(\omega_e-\omega_\uparrow-\omega)+\gamma]}\alpha. \tag{44}$$

Eq. (43) shows that when the input increases, the intensity of the cavity field also increase, resulting in the condition Eq. (15) not being satisfied.

With Eqs. (43) and (44), we can calculate Eq. (2). Suppose that the cavity and the external field are resonant, i.e., $\omega = \omega_c$, and $\Delta = \omega_e - \omega_\uparrow - \omega_c \equiv \omega_a - \omega_c$, we obtain

$$\frac{\alpha_{R,1\uparrow}}{\alpha_{i,1\uparrow}} = 1 - \frac{2\kappa_R(i\Delta+\gamma)}{(\kappa_R+\kappa_T)(i\Delta+\gamma)+g^2}. \tag{45}$$

By using Eqs. (37) and (38), we can also have Eqs. (3) and (4), which are

$$\frac{\alpha_{T,1\uparrow}}{\alpha_{i,1\uparrow}} = -\frac{2\sqrt{\kappa_R\kappa_T}(i\Delta+\gamma)}{(\kappa_R+\kappa_T)(i\Delta+\gamma)+g^2}, \tag{46}$$
$$\frac{\alpha_{S,1\uparrow}}{\alpha_{i,1\uparrow}} = \frac{2\sqrt{\kappa_R\gamma}g}{(\kappa_R+\kappa_T)(i\Delta+\gamma)+g^2}. \tag{47}$$

### C. Cattiness calculation

Taking into account all possible light losses, the final state of the system after $M$ cycles can be expressed as $|\psi_f\rangle = \frac{1}{\sqrt{2}}(|C_{0\uparrow}\rangle|\mathrm{loss}_\uparrow\rangle|\uparrow\rangle + |C_{0\downarrow}\rangle|\mathrm{loss}_\downarrow\rangle|\downarrow\rangle) = \frac{1}{\sqrt{2}}\left[\frac{1}{\sqrt{2}}(|\uparrow\rangle + |\downarrow\rangle)|\psi_0\rangle + \frac{1}{\sqrt{2}}(|\uparrow\rangle - |\downarrow\rangle)|\psi_1\rangle\right]$. After measuring the atom, the light field collapses to state $|\psi_{c=0,1}\rangle = \frac{1}{\sqrt{2}}(|C_{0\uparrow}\rangle|\mathrm{loss}_\uparrow\rangle + (-1)^c|C_{0\downarrow}\rangle|\mathrm{loss}_\downarrow\rangle)$ with equal probability. Its corresponding density matrix is denoted as $|\psi_c\rangle\langle\psi_c|$. By tracing over the loss terms, the resulting reduced density matrix is obtained, represented as

$$\rho = Tr_{\mathrm{loss}}\{|\psi_c\rangle\langle\psi_c|\} = \frac{1}{2}|C_{0\uparrow}\rangle\langle C_{0\uparrow}| + \frac{1}{2}|C_{0\downarrow}\rangle\langle C_{0\downarrow}|$$
$$+(-1)^c\frac{1}{2}\langle\mathrm{loss}_\downarrow|\mathrm{loss}_\uparrow\rangle|C_{0\uparrow}\rangle\langle C_{0\downarrow}| + (-1)^c\frac{1}{2}\langle\mathrm{loss}_\uparrow|\mathrm{loss}_\downarrow\rangle|C_{0\downarrow}\rangle\langle C_{0\uparrow}| \tag{48}$$

By substituting $\rho$ into the definition of cattiness [69-71]

$$C_a = -Tr[\rho \cdot L(\rho)] \tag{49}$$

with $L(\rho) = a\rho a^+ - \frac{1}{2}\rho a^+ a - \frac{1}{2}a^+ a\rho$, we can obtain that

$$C_a = -\sum_{v=0}^{\infty}\langle v|\rho a\rho a^+|v\rangle + \sum_{v=0}^{\infty} v\langle v|\rho^2|v\rangle \tag{50}$$

where $|v\rangle$ represents Fock state. Utilizing the following expressions ($j = \uparrow, \downarrow$),

$$\langle v|C_{0j}\rangle = \langle v|e^{-\frac{|C_{0j}|^2}{2}}\sum_{u=0}^{\infty}\frac{C_{0j}^u}{\sqrt{u!}}|u\rangle = e^{-\frac{|C_{0j}|^2}{2}}\frac{C_{0j}^v}{\sqrt{v!}} \tag{51}$$

we have

$$C_{\mathrm{a}} = \frac{1}{4}(|\langle \mathrm{loss}_{\uparrow}|\mathrm{loss}_{\downarrow}\rangle|^2 - |\langle C_{0\uparrow}|C_{0\downarrow}\rangle|^2)|C_{0\uparrow} - C_{0\downarrow}|^2 \quad (52)$$

For comparison, we also give the expression for fidelity here, which is

$$F = \frac{1}{4}\{|\langle \alpha|C_{0\uparrow}\rangle|^2 + |\langle -\alpha|C_{0\downarrow}\rangle|^2 + 2\mathrm{Re}[\langle \alpha|C_{0\uparrow}\rangle\langle C_{0\downarrow}|-\alpha\rangle\langle \mathrm{loss}_{\downarrow}|\mathrm{loss}_{\uparrow}\rangle]\} \quad (53)$$

It can be easily seen that the fidelity is related to $|\langle \mathrm{loss}_{\uparrow}|\mathrm{loss}_{\downarrow}\rangle|$, while the cattiness is related to $|\langle \mathrm{loss}_{\uparrow}|\mathrm{loss}_{\downarrow}\rangle|^2$, and thus is more sensitive to light loss.

**Data availability:**

All data generated or analyzed during the simulations of study are available from the corresponding author on reasonable request.

**Code availability:**

The code employed to generate the simulations presented in this work is available upon request. Interested parties may inquire about access to the code by contacting the corresponding author.

**Author contributions:**

The theory was conceived by Z-H.L, M.A. and M.S.Z. Numerical calculations were performed by F.Y. and Z-Y.L. under the supervision of Z-H.L. All the authors participated in the manuscript preparation, discussions, and checks of the results.

**Corresponding authors:**

Correspondence to Zheng-Hong Li or M. Al-Amri or M. Suhail Zubairy.

**Competing interests:**

The authors declare no competing interests.

| Effective fidelity | | $F_{\mathrm{ef}} \geq 70\%$ | $F_{\mathrm{ef}} \geq 80\%$ | $F_{\mathrm{ef}} \geq 90\%$ | $F_{\mathrm{ef}} \geq 95\%$ |
|---|---|---|---|---|---|
| $\lvert\alpha\rvert^2$ | $M$ | $\varepsilon$ | $\varepsilon$ | $\varepsilon$ | $\varepsilon$ |
| 3 | 6 | $\leq 3.91 \times 10^{-2}$ ($\lvert\alpha_{\mathrm{ef}}\rvert^2$ =2.36) | $\leq 1.31 \times 10^{-2}$ ($\lvert\alpha_{\mathrm{ef}}\rvert^2$ =2.77) | | |
| 3 | 10 | $\leq 2.95 \times 10^{-2}$ ($\lvert\alpha_{\mathrm{ef}}\rvert^2$ =2.22) | $\leq 1.21 \times 10^{-2}$ ($\lvert\alpha_{\mathrm{ef}}\rvert^2$ =2.66) | $1.87 \times 10^{-3}$ ($\lvert\alpha_{\mathrm{ef}}\rvert^2$ =2.94) | |
| 4 | 8 | $\leq 2.14 \times 10^{-2}$ ($\lvert\alpha_{\mathrm{ef}}\rvert^2$ =3.36) | $\leq 7.31 \times 10^{-3}$ ($\lvert\alpha_{\mathrm{ef}}\rvert^2$ =3.77) | | |
| 4 | 10 | $\leq 1.91 \times 10^{-2}$ ($\lvert\alpha_{\mathrm{ef}}\rvert^2$ =3.30) | $\leq 7.36 \times 10^{-3}$ ($\lvert\alpha_{\mathrm{ef}}\rvert^2$ =3.72) | $\leq 2.38 \times 10^{-5}$ ($\lvert\alpha_{\mathrm{ef}}\rvert^2$ =4.00) | |
| 4 | 50 | $\leq 5.18 \times 10^{-3}$ ($\lvert\alpha_{\mathrm{ef}}\rvert^2$ =3.08) | $\leq 2.53 \times 10^{-3}$ ($\lvert\alpha_{\mathrm{ef}}\rvert^2$ =3.52) | $\leq 9.17 \times 10^{-4}$ ($\lvert\alpha_{\mathrm{ef}}\rvert^2$ =3.82) | $\leq 3.05 \times 10^{-4}$ ($\lvert\alpha_{\mathrm{ef}}\rvert^2$ =3.94) |
| 8 | 20 | $\leq 4.53 \times 10^{-3}$ ($\lvert\alpha_{\mathrm{ef}}\rvert^2$ =7.31) | $\leq 1.81 \times 10^{-3}$ ($\lvert\alpha_{\mathrm{ef}}\rvert^2$ =7.71) | | |
| 8 | 50 | $\leq 2.22 \times 10^{-3}$ ($\lvert\alpha_{\mathrm{ef}}\rvert^2$ =7.16) | $\leq 1.09 \times 10^{-3}$ ($\lvert\alpha_{\mathrm{ef}}\rvert^2$ =7.58) | $\leq 3.36 \times 10^{-4}$ ($\lvert\alpha_{\mathrm{ef}}\rvert^2$ =7.87) | $\leq 3.96 \times 10^{-5}$ ($\lvert\alpha_{\mathrm{ef}}\rvert^2$ =7.98) |
| 16 | 50 | $\leq 9.47 \times 10^{-4}$ ($\lvert\alpha_{\mathrm{ef}}\rvert^2$ =15.3) | $\leq 4.18 \times 10^{-4}$ ($\lvert\alpha_{\mathrm{ef}}\rvert^2$ =15.7) | $\leq 5.50 \times 10^{-5}$ ($\lvert\alpha_{\mathrm{ef}}\rvert^2$ =16.0) | |

Table.1: Impact of classical optical devices on fidelity.

We specify the tolerance to light loss ($\varepsilon$) required by classical optical devices in a single cycle to achieve a specific effective fidelity ($F_{\mathrm{ef}}$) across various $\lvert\alpha\rvert^2$ and $M$ values. The corresponding output light intensity $\lvert\alpha_{\mathrm{ef}}\rvert^2$ is indicated in parentheses. The parameters are set as $(g, \gamma, \kappa_{\mathrm{R}}) = 2\pi \times (7.8, 3.0, 2.3)\mathrm{MHz}$ and $\kappa_{\mathrm{T}} = \Delta = 0$.

| Cattiness | | $Ca \geq 50\%\lvert\alpha\rvert^2$ | $Ca \geq 60\%\lvert\alpha\rvert^2$ | $Ca \geq 70\%\lvert\alpha\rvert^2$ | $Ca \geq 80\%\lvert\alpha\rvert^2$ |
|---|---|---|---|---|---|
| $\lvert\alpha\rvert^2$ | $M$ | $\varepsilon$ | $\varepsilon$ | $\varepsilon$ | $\varepsilon$ |
| 3 | 6 | $\leq 2.48 \times 10^{-3}$ ($\lvert\alpha_{\mathrm{ef}}\rvert^2$ =2.96) | | | |
| 3 | 10 | $\leq 4.67 \times 10^{-3}$ ($\lvert\alpha_{\mathrm{ef}}\rvert^2$ =2.86) | $\leq 2.02 \times 10^{-3}$ ($\lvert\alpha_{\mathrm{ef}}\rvert^2$ =2.94) | | |
| 4 | 8 | $\leq 1.53 \times 10^{-3}$ ($\lvert\alpha_{\mathrm{ef}}\rvert^2$ =3.95) | | | |
| 4 | 10 | $\leq 2.43 \times 10^{-3}$ ($\lvert\alpha_{\mathrm{ef}}\rvert^2$ =3.90) | $\leq 4.13 \times 10^{-4}$ ($\lvert\alpha_{\mathrm{ef}}\rvert^2$ =3.98) | | |
| 4 | 50 | $\leq 1.37 \times 10^{-3}$ ($\lvert\alpha_{\mathrm{ef}}\rvert^2$ =3.73) | $\leq 9.37 \times 10^{-4}$ ($\lvert\alpha_{\mathrm{ef}}\rvert^2$ =3.82) | $\leq 5.79 \times 10^{-4}$ ($\lvert\alpha_{\mathrm{ef}}\rvert^2$ =3.89) | $\leq 2.76 \times 10^{-4}$ ($\lvert\alpha_{\mathrm{ef}}\rvert^2$ =3.95) |
| 8 | 20 | $\leq 6.81 \times 10^{-4}$ ($\lvert\alpha_{\mathrm{ef}}\rvert^2$ =7.89) | $\leq 1.45 \times 10^{-4}$ ($\lvert\alpha_{\mathrm{ef}}\rvert^2$ =7.98) | | |
| 8 | 50 | $\leq 6.04 \times 10^{-4}$ ($\lvert\alpha_{\mathrm{ef}}\rvert^2$ =7.76) | $\leq 3.83 \times 10^{-4}$ ($\lvert\alpha_{\mathrm{ef}}\rvert^2$ =7.85) | $\leq 1.99 \times 10^{-4}$ ($\lvert\alpha_{\mathrm{ef}}\rvert^2$ =7.92) | $\leq 4.12 \times 10^{-5}$ ($\lvert\alpha_{\mathrm{ef}}\rvert^2$ =7.98) |
| 16 | 50 | $\le 1.99\times10^{-4}$ ($\lvert\alpha_{\mathrm{ef}}\rvert^2$ =15.8) | $\le 8.83\times10^{-5}$ ($\lvert\alpha_{\mathrm{ef}}\rvert^2$ =15.9) | | |

Table.2: Impact of classical optical devices on cattiness.

We specify the tolerance to light loss ($\varepsilon$) required by classical optical devices in a single cycle to achieve a specific cattiness ($C_{\mathrm{a}}$) across various $|\alpha|^2$ and $M$ values. The corresponding output light intensity $|\alpha_{\mathrm{ef}}|^2$ is indicated in parentheses. The parameters are set as $(g,\gamma,\kappa_{\mathrm{R}})=2\pi\times(7.8,3.0,2.3)\mathrm{MHz}$ and $\kappa_{\mathrm{T}}=\Delta=0$.

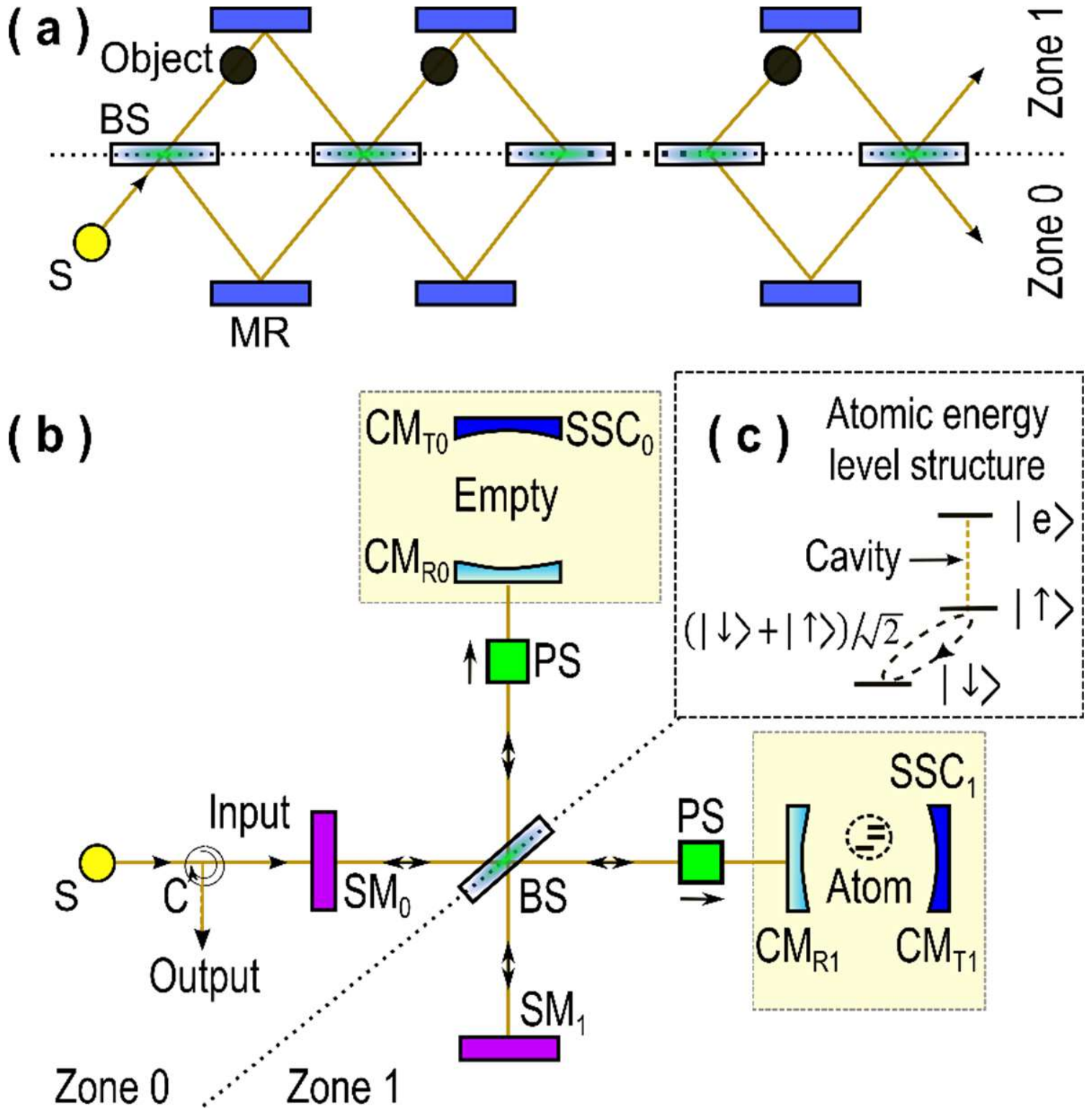

FIG. 1 Schematic for cat state preparation.

S represents the coherent light source, BS the beam-splitter, MR the normal mirror, SM the switchable mirror, PS the phase shifter, and C the optical circulator. (a) Principle model based on a chain interferometer. (b) Multiple reflection scheme based on a Michelson interferometer. When the switchable mirrors (SM) are turned on, the coherent light pulse bounces inside the interferometer and interacts with the single-side cavity (SSC) $M$ times. The SSC consists of two mirrors, $\mathrm{CM_T}$ and $\mathrm{CM_R}$, with $\mathrm{CM_T}$ assumed to have perfect reflectivity under ideal conditions. Only $\mathrm{SSC_1}$ contains a three-level atom. (c) The energy level structure of the atom, where the cavity mode couples only with the transition between levels $|\uparrow\rangle$ and $|e\rangle$. The atom is prepared in superposition state $(|\downarrow\rangle+|\uparrow\rangle)/\sqrt{2}$.

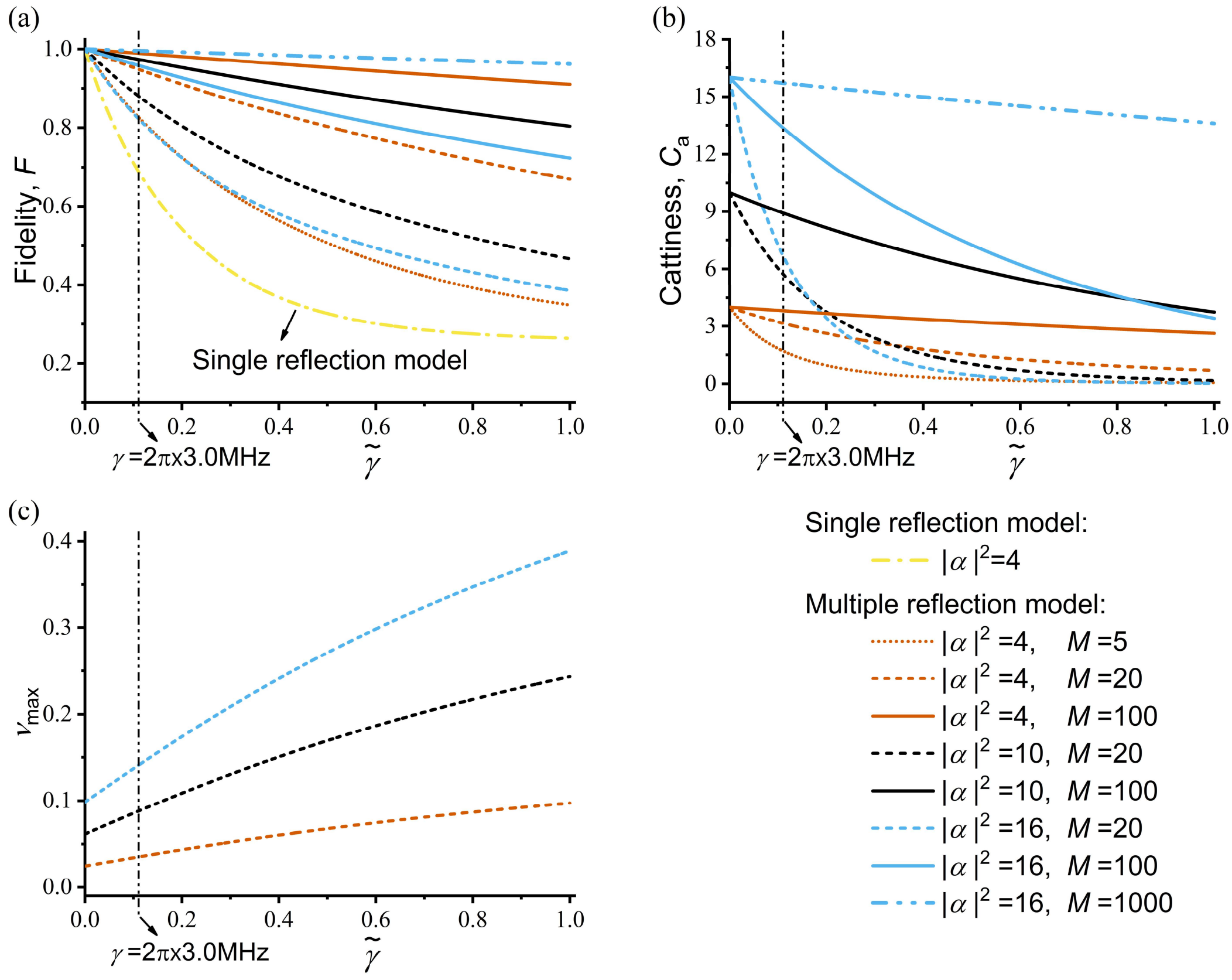

Fig.2 Influence of atomic spontaneous emission decay.

(a) Fidelity $F$ and (b) cattiness $C_{\mathrm{a}}$ versus dimensionless $\tilde{\gamma} = \kappa_{\mathrm{R}}\gamma/g^2$ with $g = 2\pi \times 7.8\mathrm{MHz}$, $\kappa_R = \kappa = 2\pi \times 2.3\mathrm{MHz}$ and $\Delta = 0$. The yellow dotted-dashed curve represents the single reflection case. Other curves represent the multiple reflection case with different colors indicating different $|\alpha|^2$ and different styles indicating different $M$. (c) We plot $v_{\max}$ versus $\tilde{\gamma}$ when $M = 20$, where $v_{\max}$ is the maximum value of the average photon number reaching $\mathrm{SSC}_1$ in each single cycle across all cycles when the atom is in $|\uparrow\rangle$.

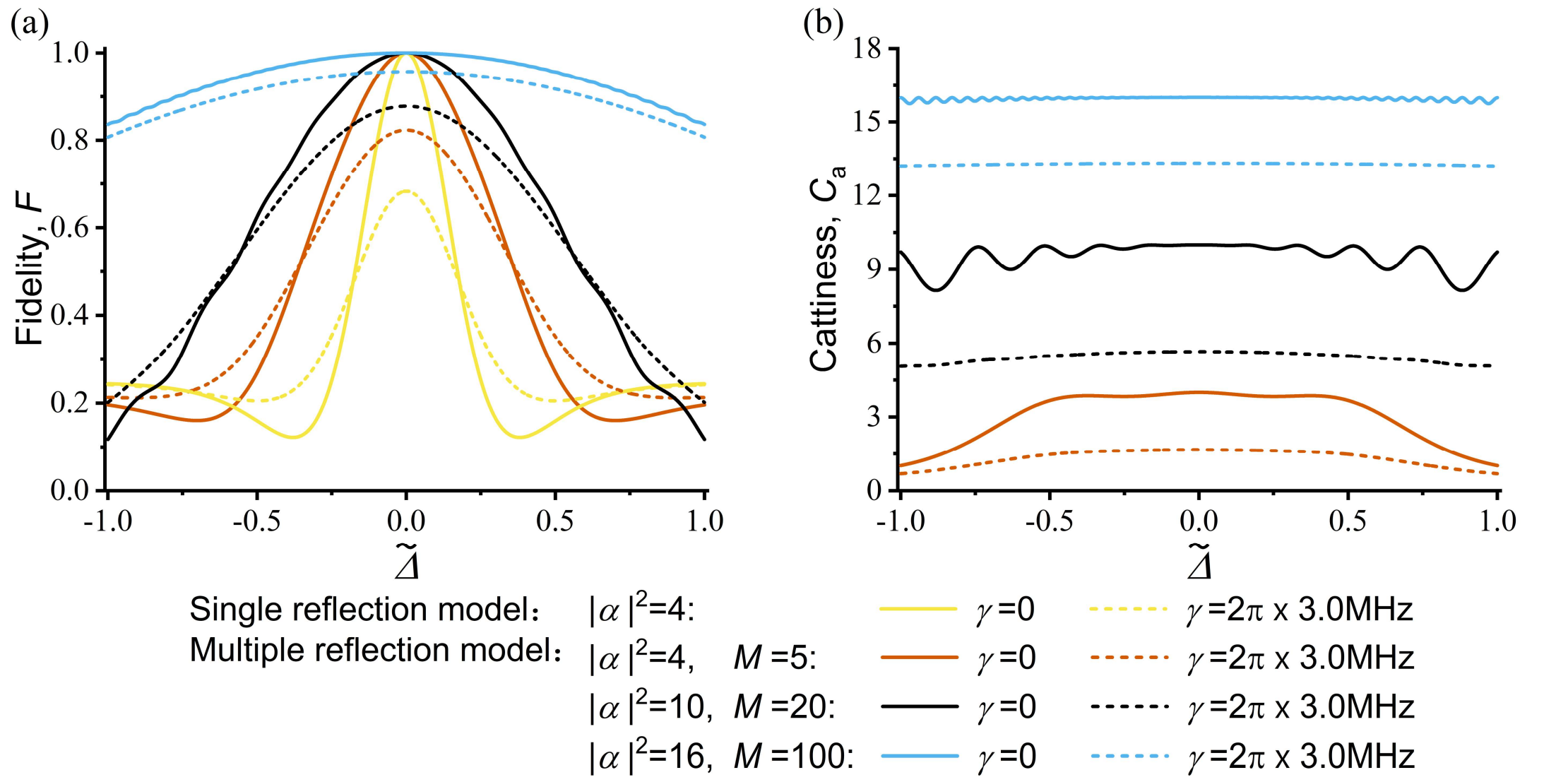

Fig.3 Influence of atom-cavity detuning.

(a) Fidelity $F$ and (b) cattiness $C_{\mathrm{a}}$ versus dimensionless $\widetilde{\Delta} = \kappa_{\mathrm{R}}\Delta/g^2$ with $g = 2\pi \times 7.8\mathrm{MHz}$ and $\kappa_{\mathrm{R}} = \kappa = 2\pi \times 2.3\mathrm{MHz}$. The solid curves are plotted for $\gamma = 0$, and dashed curves are for $\gamma = 2\pi \times 3.0\mathrm{MHz}$. The yellow curves represent the single reflection model with $|\alpha|^2 = 4$, while other curves are for the multiple reflection scheme.

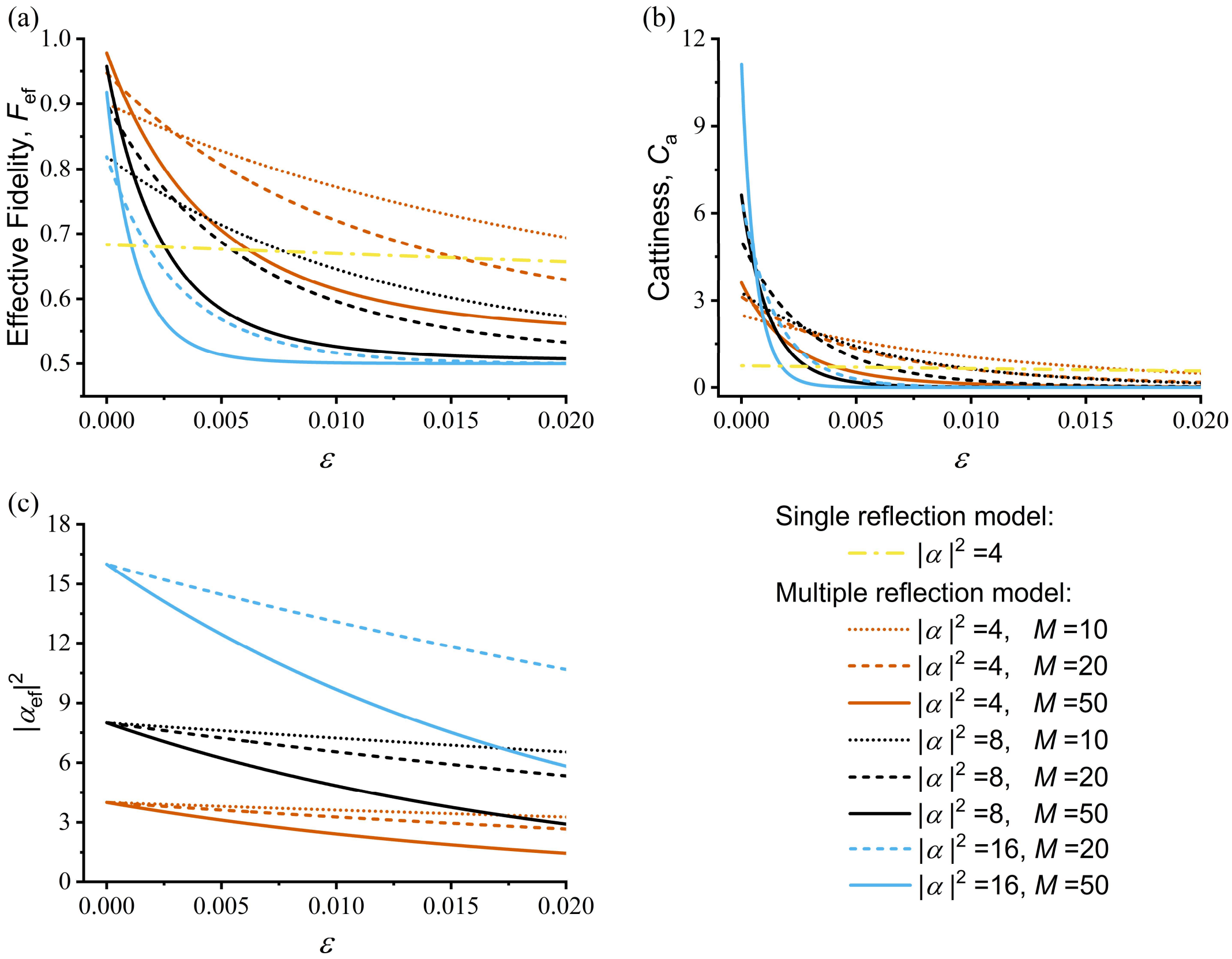

Fig.4 Influence of light loss caused by classical optical devices.
(a) Effective fidelity $F_{\rm ef}$, (b) cattiness $C_{\rm a}$, and (c) the output intensity $|\alpha_{\rm ef}|^2$ versus $\varepsilon$ with $g = 2\pi \times 7.8\text{MHz}$, $\gamma = 2\pi \times 3.0\text{MHz}$, $\kappa_{\rm R} = \kappa = 2\pi \times 2.3\text{MHz}$, and $\Delta = 0$. The yellow dotted-dashed curves represent the single reflection model with $|\alpha|^2 = 4$, and other curves represent the multiple reflection scheme with different $|\alpha|^2$ and $M$. The target state is $(|\alpha_{\rm ef}\rangle|\uparrow\rangle + |-\alpha_{\rm ef}\rangle|\downarrow\rangle)/\sqrt{2}$.

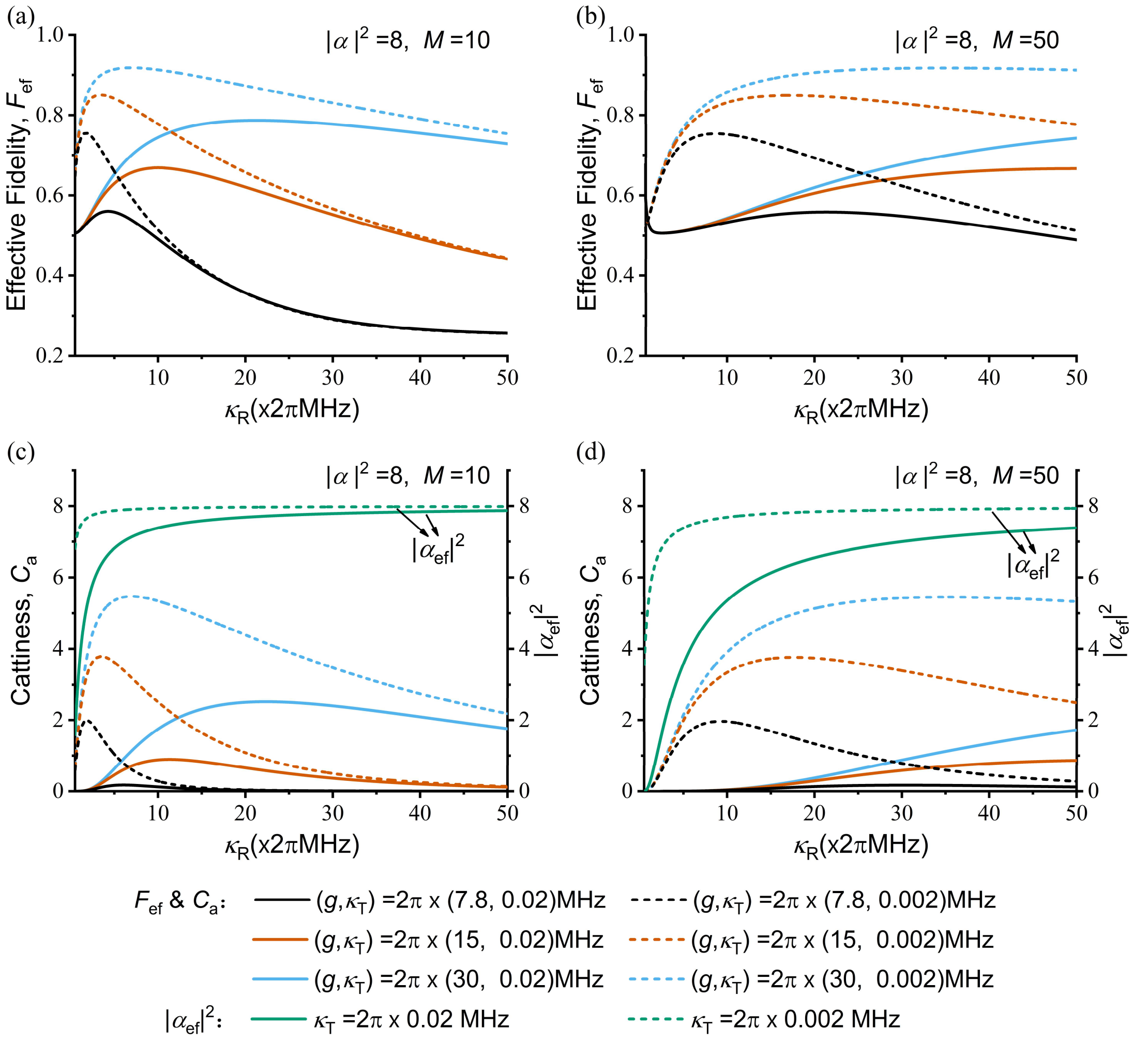

Fig.5 Influence of cavity parameters.

(a, b) Effective fidelity $F_{\rm ef}$, (c, d) cattiness $C_{\rm a}$, and the output intensity $|\alpha_{\rm ef}|^2$ versus $\kappa_{\rm R}$ with varying $g$, $M$, and $\kappa_{\rm T}$ in the multiple reflection model, where other parameters are set as $|\alpha|^2 = 8$, $\gamma = 2\pi \times 3.0\text{MHz}$, and $\Delta = 0$. The target state is $(|\alpha_{\rm ef}\rangle|\uparrow\rangle + |-\alpha_{\rm ef}\rangle|\downarrow\rangle)/\sqrt{2}$. Solid curves represent $\kappa_{\rm T} = 2\pi \times 0.02\text{MHz}$, and dashed curves represent $\kappa_{\rm T} = 2\pi \times 0.002\text{MHz}$. The green curves are plotted for $|\alpha_{\rm ef}|^2$ only, while curves in other colors represent $F_{\rm ef}$ and $C_{\rm a}$. Black corresponds to $g = 2\pi \times 7.8\text{MHz}$, red to $g = 2\pi \times 15\text{MHz}$, and blue to $g = 2\pi \times 30\text{MHz}$.